\PassOptionsToPackage{table}{xcolor}
\documentclass[acmtog]{acmart}
\usepackage{wrapfig}

\setcopyright{cc}
\setcctype{by}
\acmJournal{TOG}
\acmYear{2026} 
\acmVolume{45} 
\acmNumber{4} 
\acmArticle{65}
\acmMonth{7} 
\acmDOI{10.1145/3811315}

\usepackage{color}
\usepackage{xcolor}
\usepackage{graphicx}
\usepackage{booktabs}
\usepackage{amsmath, amsthm, amsfonts, amssymb}
\usepackage{mathrsfs}
\usepackage{textcomp}
\usepackage{epstopdf}
\usepackage{multirow}
\usepackage{wrapfig}
\usepackage{subfig}
\usepackage{booktabs} % For formal tables
\usepackage[ruled,linesnumbered]{algorithm2e} % For algorithms
\usepackage{algorithmicx}  
\usepackage{algpseudocode}
\usepackage{ragged2e}
\usepackage[normalem]{ulem}
\usepackage{cleveref}
\usepackage{bm}
\usepackage{appendix}
\usepackage{enumitem}

\newcommand{\nosemic}{\renewcommand{\@endalgocfline}{\relax}}% Drop semi-colon ;
\newcommand{\dosemic}{\renewcommand{\@endalgocfline}{\algocf@endline}}% Reinstate semi-colon ;
\let\oldnl\nl% Store \nl in \oldnl
\newcommand{\nonl}{\renewcommand{\nl}{\let\nl\oldnl}}% Remove line number for one line

\SetAlFnt{\small}
\SetAlCapFnt{\small}
\SetAlCapNameFnt{\small}
\SetAlCapHSkip{0pt}

\definecolor{green}{rgb}{0, 0.5, 0}
\definecolor{orange}{rgb}{0.6, 0.3, 0.1}
\definecolor{red}{rgb}{1.0, 0.0, 0.0}
\definecolor{teal}{rgb}{0.0, 0.4, 0.4}
\definecolor{purple}{rgb}{0.65,0,0.65}
\definecolor{saffron}{rgb}{0.95,0.75,0.2}
\definecolor{lightblue}{rgb}{0.0,0.5,1}
\definecolor{brown}{rgb}{0.5, 0.16, 0.16}
\definecolor{brickred}{rgb}{.6, .2 .1}
\definecolor{coral}{rgb}{1,0.45,0.33}
\definecolor{newcolor}{rgb}{.8,.349,.1}

\newcommand{\re}[1]{{\color{black}#1}}
\newcommand{\rre}[1]{{\color{black}#1}}

\begin{document}

\title{Fast VEM Fluid Simulation}

\author{Runze Zhang}
\email{oliverzrz.cyber@gmail.com}
\orcid{0009-0008-6304-623X}
\affiliation{%
  \department{VCIP, College of Computer Science}
  \institution{Nankai University}
   \city{TianJin} 
  \country{China}}

\author{Bo Ren}
\authornote{Corresponding author: Bo Ren (rb@nankai.edu.cn)}
\email{rb@nankai.edu.cn}
\orcid{0000-0001-8179-9122}
\affiliation{
  \department{VCIP, College of Computer Science}
  \institution{Nankai University}
  \city{TianJin} 
  \country{China}}

\renewcommand{\shortauthors}{Runze Zhang, Bo Ren}

\begin{abstract}

The intricate motion arising from fluid--boundary interactions is visually compelling, yet notoriously difficult and computationally expensive to simulate in the presence of complex boundaries.
Accurately resolving boundary geometry requires body-fitted
grids constructed via cut-cell methods, which often leads to poorly conditioned
linear systems in the pressure projection stage and, consequently, prohibitive
computational cost.
We present \textit{FastVEM}, an efficient boundary-conforming fluid simulation
framework that enables high-fidelity flow--boundary interaction at substantially
reduced cost. Computational efficiency is achieved through a coordinated,
top-down design spanning numerical discretization, grid construction, and linear
solvers.  FastVEM adopts a Virtual Element Method (VEM) discretization to robustly
enforce incompressibility and boundary conditions on irregular body-fitted grids,
and employs a VEM
polynomial-space Particle-in-Cell scheme for advection.
Complementing this discretization, a convexity-preserving cut-cell strategy is
introduced to construct simulation-friendly body-fitted grids. To accelerate pressure projection, we develop a Galerkin geometric multigrid solver
featuring a diffusion-free prolongation operator that prevents coarse-level
matrix densification, along with a nested, boundary-aware grid hierarchy that
ensures well-posed placement of coarse-level degrees of freedom.
Compared to prior cut-cell--based fluid simulators, \rre{FastVEM speeds up the computationally dominant pressure projection stage by up to 100$\times$},
while robustly handling even more challenging boundary geometries.

\end{abstract}



% \keywords{Fluid Dynamics, Virtual Element Method, Cut-Cell, Multigrid Methods}

\begin{teaserfigure}
  \centering
  \includegraphics[width=\linewidth]{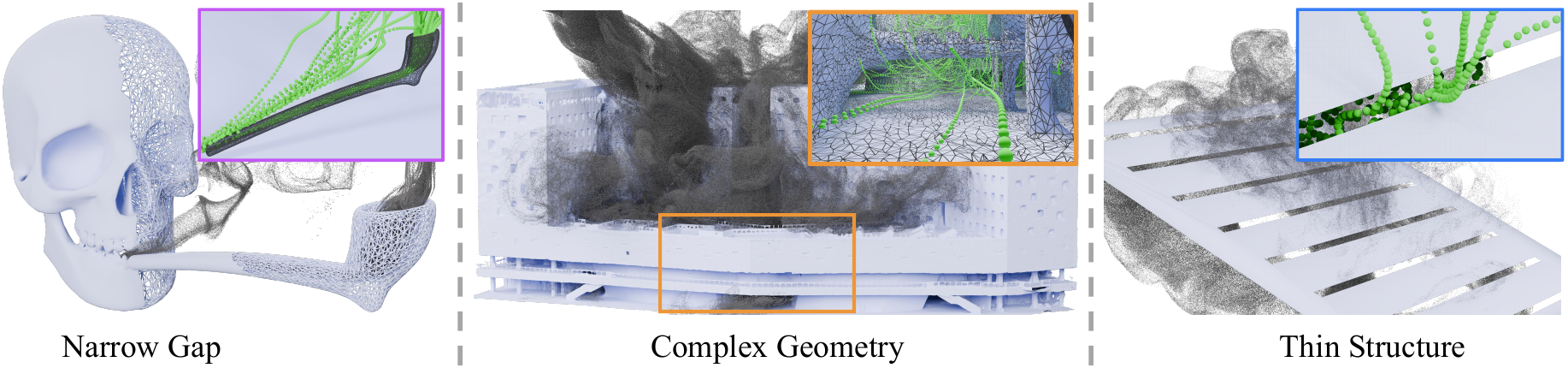}
  \caption{FastVEM provides an efficient boundary-conforming fluid simulation framework
that robustly handles narrow gaps (left), complex geometry reconstructed from
drone scans (middle), and ultra-thin sheet boundary with a relative thickness of $10^{-5}$
(right). Embedding these challenging geometries into a $128^3$ Cartesian grid,
high-fidelity smoke simulation using a second-order VEM discretization for the pressure projection requires
less than $1\,\mathrm{min}$/frame. Smoke trajectories are visualized as green
particles in the zoomed-in views;  the middle inset highlights smoke motion near
the interior base of the building.
}
\label{fig:teaser}
\end{teaserfigure}
\maketitle
\section{Introduction}
\label{sec:intro}
Ensuring that fluid motion conforms to geometrically complex boundaries is essential
for producing physically plausible animations. Despite substantial progress, designing 
an efficient boundary-conforming fluid simulator remains challenging for two primary
reasons. First, Cartesian grid--based simulators achieve high computational efficiency, but struggle with non--grid-aligned boundaries. Even when augmented with area-fraction techniques, such methods remain limited in
accurately capturing complex boundary geometry within a single grid cell. Increasing the grid resolution can mitigate this limitation, but quickly leads to impractical memory and computational costs. Second, body-fitted grid--based fluid simulation enables accurate capture of complex
geometries. However, the resulting irregular body-fitted grids often lead to poorly conditioned linear systems, posing a significant challenge for
the efficient enforcement of incompressibility and boundary conditions.

Existing works on boundary-conforming fluid simulation generally follow two  directions.
Adaptive resolution methods~\cite{octree03,octree20,powd_octree,whirp} employ local grid refinement near boundaries to avoid a global increase in grid resolution.
While such approaches capture finer boundary features at reduced cost, approximating arbitrary boundary surfaces with axis-aligned grids remains inherently inflexible: accurately resolving even a single non-axis-aligned plane can trigger aggressive local refinement and thus a substantial increase in the number of degrees of freedom.
\re{Alternatively, cut-cell--based methods address this limitation by simulating fluids on body-fitted grids that conform directly to boundaries~\cite{vempic,batty16,batty17,extended_cut_cell}.
% These methods offer higher geometric fidelity; however, enforcing incompressibility and boundary conditions on such irregular grids requires solving poorly conditioned pressure Poisson systems, leading to degraded accuracy and prohibitive computational cost.
These methods offer higher geometric fidelity; however, accurately resolving fine-scale structures often leads to irregular grids with poorly shaped cut cells, thereby increasing the cost of enforcing incompressibility and boundary conditions.}
 Efficient multigrid solvers can alleviate this issue; however, constructing
multigrid methods on body-fitted grids remains challenging. Non--Galerkin
approaches struggle to ensure consistent boundary condition transfer
across multigrid levels, while Galerkin projection on irregular grids tends to
induce connectivity proliferation among degrees of freedom, leading to dense
coarse--level system matrices~\cite{Tiantian}.

Following the cut-cell paradigm, we propose \textit{FastVEM}, an efficient
boundary-conforming fluid simulation framework. FastVEM adopts a carefully coordinated, top-down design spanning numerical
discretization, grid construction, and linear system solvers. 
\re{At the numerical level, FastVEM employs a second-order Virtual Element Method (i.e., VEM of order k = 2) for pressure discretization to accurately enforce boundary conditions on
body-fitted grids.}
For grid construction, a convexity-preserving, binary space partition--based
cut-cell strategy is introduced to mitigate poorly conditioned system matrices
caused by non--star-shaped elements. Finally, pressure projection is accelerated by a geometric multigrid solver
featuring a diffusion-free prolongation operator to prevent coarse-level matrix
densification, together with a nested, boundary-aware grid hierarchy that
ensures well-posed placement of coarse-level degrees of freedom. 
Together, VEM discretization delivers the accuracy and robustness needed for boundary handling, while the proposed cut-cell strategy and multigrid
method alleviate the otherwise prohibitive computational cost of 
VEM schemes. With this balance, FastVEM robustly supports a wide range of  boundaries and
achieves up to two orders of magnitude speedup over prior cut-cell--based fluid
solvers in the computationally dominant pressure projection stage.

FastVEM extends the capabilities of cut-cell fluid simulators, making boundary-conforming fluid simulation more affordable. Algorithm~\ref{alg:fastvem} provides an overview of the proposed framework, and the main contributions are summarized as follows: 
\begin{itemize}[topsep=3pt,itemsep=2pt,parsep=0pt]
    \item A VEM-based cut-cell fluid simulator with second-order pressure discretization
    for accurate enforcement of boundary conditions, \rre{speeding up the computationally dominant pressure projection stage by up to 100$\times$ over existing cut-cell--based fluid simulators.}
    %  achieving more than
    % 100$\times$ speedup over existing cut-cell--based solvers at the primary
    % computational bottleneck, the pressure projection stage.
    \item A geometric multigrid method for VEM Poisson problems uses a Galerkin formulation to ensure consistent transfer of boundary conditions across levels, and a diffusion-free prolongation operator to avoid coarse-level system matrix densification caused by Galerkin projection.

    % \item A geometric multigrid method for VEM Poisson problems, featuring a diffusion-free prolongation operator that improves efficiency by avoiding coarse-level matrix densification.    
    \item A binary space partitioning–based cut-cell strategy for constructing simulation-friendly body-fitted grids, together with a nested, boundary-aware grid hierarchy construction strategy that supports the proposed multigrid method.
\end{itemize}

\re{The code for this paper is located at \url{https://oliver-zrz-cyber.github.io/FastVEM/}.}

\section{Related Work}
\label{sec:rw}

\noindent\textbf{Boundary-Conforming Fluid Simulation.}
Accurate conformity between fluid motion and complex immersed geometry is crucial
for physically plausible flows. Existing boundary-conforming fluid solvers can be
broadly categorized into three classes. The first class of methods enhances boundary fidelity by locally adapting grid
resolution on top of a Cartesian grid. Representative approaches include
adaptive octree structures that resolve fine-scale details~\cite{octree03},
power-diagram--based methods that avoid T-junction artifacts while achieving
second-order accurate pressure projection near boundaries~\cite{powd_octree},
and surface-adaptive octree frameworks designed for efficient and robust liquid
simulation~\cite{octree20}. Within this category, other variants increase local
resolution by overlaying high-resolution grids on a coarse background~\cite{whirp},
or by coupling boundary-guided multiresolution grids with parametric boundary
treatments to better resolve turbulent flows near solid surfaces~\cite{plbm}.
More recently, overset grid techniques have been extended to support moving,
multi-resolution domains with explicit boundary layer control, enabling more
accurate two-phase fluid--rigid interactions~\cite{overset}. The second class of methods preserves the underlying Cartesian grid and
incorporates boundary information directly into the pressure solve by modifying
the discretization stencils used to assemble the system matrix. Representative
examples include ghost-fluid formulations that achieve second-order accurate
free-surface pressure conditions by altering finite-difference gradient
stencils~\cite{ghost}, as well as variational approaches that handle irregular
boundaries through divergence stencils augmented with volume fractions~\cite{batty07}.
Subsequent improvements replace volume fractions with area fractions in
finite-volume cut-cell discretizations to achieve second-order pressure
accuracy~\cite{areafractions}, and further improve efficiency by integrating
multigrid solvers~\cite{icutmg}. Despite their simplicity and compatibility with
Cartesian grids, these methods remain limited in accurately resolving complex
boundary geometries. The third class of methods comprises cut-cell--based fluid solvers, which
construct body-fitted grids by explicitly cutting a Cartesian grid with a
boundary mesh, thereby enabling discretizations that naturally conform to
complex geometry. Early work employs fixed tetrahedral meshes near boundaries to
improve geometric fidelity~\cite{batty10}. Subsequent approaches introduce
adaptive discontinuous Galerkin schemes on body-fitted grids to capture detailed
free-surface dynamics even on coarse grids~\cite{edwards14}. Finite-volume
cut-cell formulations further advance this line of work by improving pressure
projection and velocity interpolation near thin boundaries~\cite{batty16}, as
well as enabling robust two-way coupling through positive-definite system
construction~\cite{batty17}. \re{By assigning multiple virtual pressure samples per cell with a single physical degree of freedom, this formulation extends to support sub-grid free-surface structures, achieving second-order accuracy for pressure values while preserving the symmetric positive definite structure~\cite{extended_cut_cell}.} Related ideas have also been explored in the lattice
Boltzmann framework, where sample-based cut-cell techniques are used to resolve
sub-voxel geometric features~\cite{lbm21}. 
More recent work, VEMPIC~\cite{vempic}, combines Mandoline~\cite{mandoline} for constructing body-fitted grids with a non-conforming virtual element method for pressure projection, enabling robust fluid interaction with complex boundaries at a high computational cost. To enforce incompressibility and boundary conditions, our work likewise employs the virtual element method, but adopts a conforming formulation to avoid discontinuities in the pressure gradient across cell interfaces that are inherent to non-conforming schemes.

\noindent\textbf{Multigrid Methods.}
Enforcing incompressibility entails solving large linear systems, which typically dominate the computational cost of split-based fluid solvers. To mitigate this bottleneck, multigrid methods are widely adopted and have a long history of efficiently solving the Poisson equations arising in fluid simulation, initially on regular Cartesian grids via geometric multigrid preconditioning of conjugate gradient solvers~\cite{McAdams}.
This line of work was subsequently extended to support irregular boundaries by accelerating volume-fraction--based variational formulations~\cite{Chentanez2012}. Later efforts adapted multigrid solvers to hybrid Cartesian discretizations, including tall-grid representations~\cite{tallgrid} and sparsely populated grids with octree-style adaptive refinement~\cite{SPGrid}. More specialized developments further tailor multigrid techniques to fraction-based discretizations~\cite{icutmg}, introduce topology-aware coarsening to improve convergence in the presence of thin solid features~\cite{Dick2016}, or design dedicated smoothers for efficiently solving viscosity systems~\cite{Aanjaneya}, \re{and introduce adaptive smoothers to enable multiresolution adaptive grids for pressure correction in two-phase flows with large density contrasts~\cite{Phase-Field-FLIP}.}
 Despite these advances, existing geometric multigrid methods are not designed to operate directly on body-fitted grids. In contrast, algebraic multigrid methods make no assumptions about geometric structure, allowing them to accelerate sparse linear systems arising from discretizations on body-fitted grids~\cite{Zarifi, amgfluid, AMGCL}.
 In the context of elastic simulation, Galerkin multigrid methods have been widely applied to tetrahedral mesh--based discretizations~\cite{MGTT}. For example, Tiantian et al.~\cite{Tiantian} define
prolongation operators using piecewise constant interpolation to avoid dense
coarse-level matrices on tetrahedral meshes, while Liu et al.~\cite{LJM} combine the
Full Approximation Scheme with block Jacobi smoothers to construct efficient
multigrid solvers for such discretizations. Despite their success, these methods
cannot be directly applied to body-fitted fluid simulation. Beyond graphics, two works are most closely related to ours. The first proposes a $p$-multigrid method for VEM discretizations~\cite{vempmg}, performing coarse-grid correction by switching to lower-order VEM spaces rather than coarsening the volumetric grid. The second develops a geometric multigrid framework for first-order two-dimensional VEM~\cite{vemgmg}, defining the prolongation operator directly from the VEM degrees of freedom and providing theoretical convergence guarantees. To the best of our knowledge, we present the first geometric multigrid method for second-order VEM in three dimensions.

\noindent\textbf{Numerical Discretization on Irregular Grids.} 
Irregular grids offer a flexible representation for complex boundary geometries,
motivating extensive research on solving PDEs over arbitrarily shaped cells. Finite volume methods~\cite{fvm} naturally extend to general polyhedral grids, but often involve complex stencils and may lose symmetry or positive definiteness, leading to reduced accuracy on highly irregular grids.
Discontinuity-based methods, such as Discontinuous Galerkin~\cite{HDG}, Weak Galerkin~\cite{WG}, and Hybrid High-Order methods~\cite{HHO}, relax inter-element continuity and enable robust discretizations on general polyhedral grids through the use of discontinuous skeletal structures.
 Polyhedral finite element
methods~\cite{PFM} generalize classical finite elements to polygonal and
polyhedral cells, but rely on explicit rational basis functions whose evaluation
requires high-order numerical quadrature, resulting in significant computational
cost.
The Mimetic Finite Difference (MFD) method~\cite{MFD} bridges finite differences and
finite elements by constructing discrete variational formulations directly on
degrees of freedom, enabling stable and consistent discretizations on general
grids without explicit shape functions. Building on this mimetic philosophy, the
Virtual Element Method (VEM)~\cite{vemtheory} introduces a Galerkin framework with
well-defined approximation spaces and projector-based bilinear forms, overcoming
key limitations of MFD in nonlinear problems and theoretical analysis while
retaining flexibility on general polyhedral grids.

\section{VEMFLUID}
\label{sec:vemfluid}

\begin{algorithm}[t]
\caption{FastVEM}\label{alg:fastvem}

\nonl\textbf{Precomputation stage:}

Construct a body-fitted grid $\mathcal{P}^\ell$ from the given boundary mesh
(Section~\ref{sec:cutcell})\;

Build a grid hierarchy $\{\mathcal{P}^0, \ldots, \mathcal{P}^{\ell-1}\}$ for the proposed multigrid
method (Section~\ref{sec:multilevelgrid})\;

Assemble the prolongation operators $\mathbf{P}$ and construct the multilevel
system matrices via
$\mathbf{A}_c = \mathbf{R}\,\mathbf{A}_f\,\mathbf{P}$
(Section~\ref{sec:multigrid})\;

\nonl\textbf{A simulation step:}

\textit{Particle-to-grid transfer:} Velocity degrees of freedom at vertices of $P^l$ are computed by accumulating
contributions from particles in the local neighborhood
(Section~\ref{sec:p2g})\;

\textit{VEM-based pressure projection:} Assemble and solve the VEM Poisson system to obtain the pressure field
(Section~\ref{sec:vemfluid:projection})\;

\textit{Grid-to-particle transfer and particle advection:} Apply the pressure field to the polynomial velocity space, update particle
velocities, and advect particles
(Section~\ref{sec:g2p})\;
\end{algorithm}

We approximately solve the incompressible, inviscid fluid governed by the following Euler equations:

\begin{equation}
\begin{alignedat}{2}
\frac{\partial \mathbf{u}}{\partial t} + (\mathbf{u}\cdot\nabla)\mathbf{u} + \nabla p &= \mathbf{f} \qquad && \text{in } \Omega,\\
\nabla\cdot \mathbf{u} &= 0 \qquad && \text{in } \Omega,\\
\mathbf{u}\cdot \mathbf{n} &= 0 \qquad && \text{on } \partial\Omega,
\end{alignedat}
\end{equation}
\noindent
where $\mathbf{u}$ denotes the velocity field, $p$ the pressure, $\mathbf{f}$ the external force,
$\mathbf{n}$ the outward unit normal on $\partial\Omega$, and $\Omega$ the fluid domain.
The fluid density is set to one. 
Enforcing incompressibility and boundary conditions on body-fitted grids requires a discretization that remains stable and consistent over general polyhedral grids. We employ a second-order Virtual Element Method (VEM) for pressure projection, as it provides a theoretically grounded and practically mature framework for discretizing elliptic operators on such grids~\cite{vemtheory}.
\re{In contrast to~\cite{vempic}, we adopt a conforming formulation to avoid inter-cell discontinuities and to facilitate efficient multigrid solver design.
Conforming and non-conforming VEM differ fundamentally in their degrees of freedom (DOFs): the former enforces inter-element continuity through nodal, edge, face, and volume DOFs, yielding globally continuous approximations, whereas the latter relies on face- and volume-based moment DOFs that only weakly enforce continuity without guaranteeing pointwise consistency.
This difference allows conforming VEM to avoid inter-cell discontinuities in the pressure field and to achieve more consistent velocity behavior across cell interfaces.
As shown in Section~\ref{sec:multigrid}, conforming VEM also yields a DOF structure that supports the design of efficient diffusion-free multigrid solvers.
For a detailed discussion of conforming and non-conforming VEM, see~\cite{c_noc}.} The material derivative $D\mathbf{u}/Dt$ is discretized using a standard FLIP
formulation to reduce numerical dissipation, in which velocities are stored on
particles and advected through a grid-based velocity field. 
Throughout the paper, for a scalar field $f$, we use $f_h$ to denote its virtual
element approximation and $\vec f$ to represent the vector of Lagrange-type
coefficients associated with $f_h$ at DOFs. The same notation applies analogously to vector fields.

\subsection{VEM-Based Pressure Projection Preliminaries}
\label{sec:vemfluid:projection}
To keep the paper self-contained, this section briefly summarizes the VEM
formulation used for pressure projection; more detailed discussions can be
found in~\cite{vembook}. 

The pressure projection can be formulated as a Poisson
problem with Neumann boundary conditions:
\begin{equation}
\begin{alignedat}{2}
\Delta p &= \frac{1}{\Delta t}\,\nabla \cdot \mathbf{u}' \qquad && \text{in } \Omega,\\
\frac{\partial p}{\partial \mathbf{n}} &= \frac{1}{\Delta t}\,\mathbf{u}' \cdot \mathbf{n} \qquad && \text{on } \partial\Omega.
\end{alignedat}
\end{equation}
\noindent
where $\mathbf{u}'$ denotes the intermediate velocity field.  
The corresponding variational formulation is: find \( p \) such that
\begin{equation}
\int_{\Omega} \nabla p \cdot \nabla v \, \mathrm{d}x = \frac{1}{\Delta t} \int_{\Omega} \mathbf{u}' \cdot \nabla v \, \mathrm{d}x, \qquad \forall v \in V,
\end{equation}
where \( V \) is a suitable function space. 
Let \( V_k \) be the virtual element space of order \( k \), spanned by the basis functions \( \{\varphi_i^k\}_{i=1}^{N_{\mathrm{dof}}^k} \), which are implicitly defined through the degrees of freedom (DOFs) \( \{D_i^k\} \) by the duality condition:
\begin{equation}
D_j^k(\varphi_i^k) = \delta_{ij}, \qquad i,j = 1, \dots, N_{\mathrm{dof}}^k.
\end{equation}
\noindent
Here, $N_{\mathrm{dof}}^k$ denotes the total number of DOFs associated with the virtual element space $V_k$.
As a consequence, any function $v_h^k \in V_k$ admits a Lagrange-type interpolation representation,
\begin{equation}
v_h^k = \sum_{i=1}^{N_{\mathrm{dof}}^k} D_i^k(v_h^k)\,\varphi_i^k .
\end{equation}
\noindent
When no ambiguity arises, we omit the superscript $k$ indicating the order of the virtual element space. To more accurately capture the interaction between the fluid and the boundary, FastVEM employs second-order pressure to enforce both the boundary conditions and the incompressibility of a linear velocity field.
This leads us to seek the pressure solution in the \( V_2 \) space, resulting in the following discrete problem: find \( p_h \in V_2 \) such that
\begin{equation}
\int_{\Omega} \nabla p_h \cdot \nabla \varphi_i \, \mathrm{d}x
=
\frac{1}{\Delta t}\int_{\Omega} \mathbf{u}'_h \cdot \nabla \varphi_i \, \mathrm{d}x,
\qquad i=1,\dots,N_{\mathrm{dof}},
\end{equation}
where \( \{\varphi_i\}_{i=1}^{N_{\mathrm{dof}}} \) are the basis functions spanning \( V_2 \).

To solve the discrete variational problem, the global stiffness matrix \( K \) and the right-hand-side vector \( \vec{r} \) need to be computed:
\begin{equation}
K_{ij} = \int_\Omega \nabla \varphi_i \cdot \nabla \varphi_j \, \mathrm{d}x, \qquad i,j = 1, \dots, N_{\mathrm{dof}},
\end{equation}
\begin{equation}
\vec{r}_i = \frac{1}{\Delta t} \int_\Omega \mathbf{u}'_h \cdot \nabla \varphi_i \, \mathrm{d}x, \qquad i = 1, \dots, N_{\mathrm{dof}}.
\end{equation}
The global stiffness matrix \( K \) is assembled from the element-wise stiffness matrices \( K^E \), and the right-hand-side vector \( \vec{r} \) is computed as described in Section~\ref{sec:p2g}. Once \( K \) and \( \vec{r} \) are assembled, the DOF vector \( \vec{p} \) of
the discrete pressure field \( p_h \) is obtained by solving
\begin{equation}
K \, \vec{p} = \vec{r}.
\end{equation}

% To solve the discrete variational problem, we assemble a global stiffness matrix and a right-hand-side vector.  
% The global stiffness matrix $K$, associated with the Laplace operator, is defined by
% \begin{equation}
% K_{ij}=\int_\Omega \nabla \varphi_i \cdot \nabla \varphi_j\,\mathrm{d}x,
% \qquad i,j=1,\dots,N_{\mathrm{dof}} .
% \end{equation}
% \noindent
% In practice, $K$ is assembled from element-wise contributions. For each polyhedral cell $E \subset \Omega$, the local stiffness matrix $K^E$ is given by
% \begin{equation}
% K^E_{ij}=\int_E \nabla \varphi_i \cdot \nabla \varphi_j\,\mathrm{d}x,
% \qquad i,j=1,\dots,N_{\mathrm{dof}} .
% \end{equation}
% \noindent
% The global matrix is obtained by summing all local contributions,
% \begin{equation}
%      \label{K_assemble}
% K_{ij}=\sum_{E\subset\Omega} K^E_{ij}.
% \end{equation}
% \noindent
% The right-hand-side vector $\mathbf r$, with entries
% \begin{equation}
% \vec r_i =\frac{1}{\Delta t}\int_\Omega \mathbf u' \cdot \nabla \varphi_i\,\mathrm{d}x,
% \qquad i=1,\dots,N_{\mathrm{dof}} ,
% \end{equation}
% \noindent
% is computed as described in Section~\ref{sec:p2g}.  
% The resulting pressure DOFs $\vec p$ of $p_h$ are obtained by solving
% \begin{equation}
% K\,\vec p=\vec r.
% \end{equation}

In the finite element method (FEM), the element stiffness matrix $K^E$ is computed by integrating explicitly defined polynomial basis functions over each element.  
VEM extends FEM to support general polyhedral grids by introducing non-polynomial functions in the local virtual element spaces. As a consequence, the basis functions cannot be constructed explicitly. Rather than evaluating the stiffness matrix via explicitly constructed basis functions,
VEM is formulated so that, for any functions
$v_h, u_h \in V_k$,
the bilinear form
$\int_E \nabla v_h \cdot \nabla u_h \,\mathrm{d}x$
can be computed directly from the DOFs of $v_h$ and $u_h$.

The key idea behind computing $K^E_{ij}$ in VEM is to project virtual basis functions onto a polynomial space. Specifically, for a polyhedral cell $E$, let $V_k(E)$ denote the local virtual element space of order $k$ on $E$. VEM defines the projection operator
$\Pi_k^{\nabla}: V_k(E)\rightarrow \mathbb{P}_k(E)\subset V_k(E)$,
where $\mathbb{P}_k(E)$ is the space of three-dimensional polynomials of degree at most $k$ on each element $E$. The operator $\Pi_k^{\nabla}$ is defined element-wise for any $v_h\in V_k(E)$ by the orthogonality conditions
\begin{equation}
     \label{orth_condition}
\int_E \nabla p \cdot \nabla\bigl(\Pi_k^{\nabla} v_h - v_h\bigr)\,\mathrm{d}x = 0,
\qquad \forall\, p\in \mathbb{P}_k(E),
\end{equation}
along with the additional constraint
\begin{equation}
     \label{constraint}
P_0\!\left(\Pi_k^{\nabla} v_h - v_h\right)=0.
\end{equation}
\noindent
The operator $P_0$ fixes the null space of the orthogonality conditions and is defined as
\begin{equation}
P_0 v_h :=
\begin{cases}
\dfrac{1}{N_D}\displaystyle\sum_{i=1}^{N_D} D_i(v_h), & k=1, \\[6pt]
\dfrac{1}{|E|}\displaystyle\int_E v_h\,\mathrm{d}x, & k\ge 2,
\end{cases}
\end{equation}
\noindent
where $D_i(v_h)$ denotes the $i$-th degree of freedom of $v_h$. 
By definition of the projection, \( \Pi_k^{\nabla} v_h \) admits a polynomial expansion
\begin{equation}
     \label{polynomial_expand}
\Pi^{\nabla}_{k} v_h = \sum_{\alpha=1}^{N_k} s^h_{\alpha}\, m_{\alpha},
\end{equation}
\noindent
where $\{m_{\alpha}\}$ is a set of three-dimensional polynomials of degree less than or equal to $k$, and $N_{k}=\dim \mathbb{P}_{k}(E)$. The construction of the polynomial basis $\{m_{\alpha}\}$ is detailed in
Appendix~\ref{appendix:polynomial}.

Projecting the virtual basis functions onto the polynomial space amounts to
computing the coefficients $s^h_{\alpha}$ in Eq.~\ref{polynomial_expand}.
Substituting this expansion into the orthogonality conditions in Eq.~\ref{orth_condition} and the constraint in Eq.~\ref{constraint} yields the following linear
system:
\begin{equation}
     \label{eq:VirtualProj}
\begin{alignedat}{2}
\sum_{\alpha=1}^{N_k} s^h_{\alpha}
\int_E \nabla m_{\alpha}\cdot\nabla m_{\beta}\,\mathrm{d}x
&=
\int_E \nabla v_h\cdot\nabla m_{\beta}\,\mathrm{d}x,
\quad \beta = 1,\dots,N_k, \\
\sum_{\alpha=1}^{N_k} s^h_{\alpha}\, P_0(m_{\alpha})
&= P_0(v_h).
\end{alignedat}
\end{equation}

\begin{figure}[t!]
    \centering
    \includegraphics[width=\linewidth]{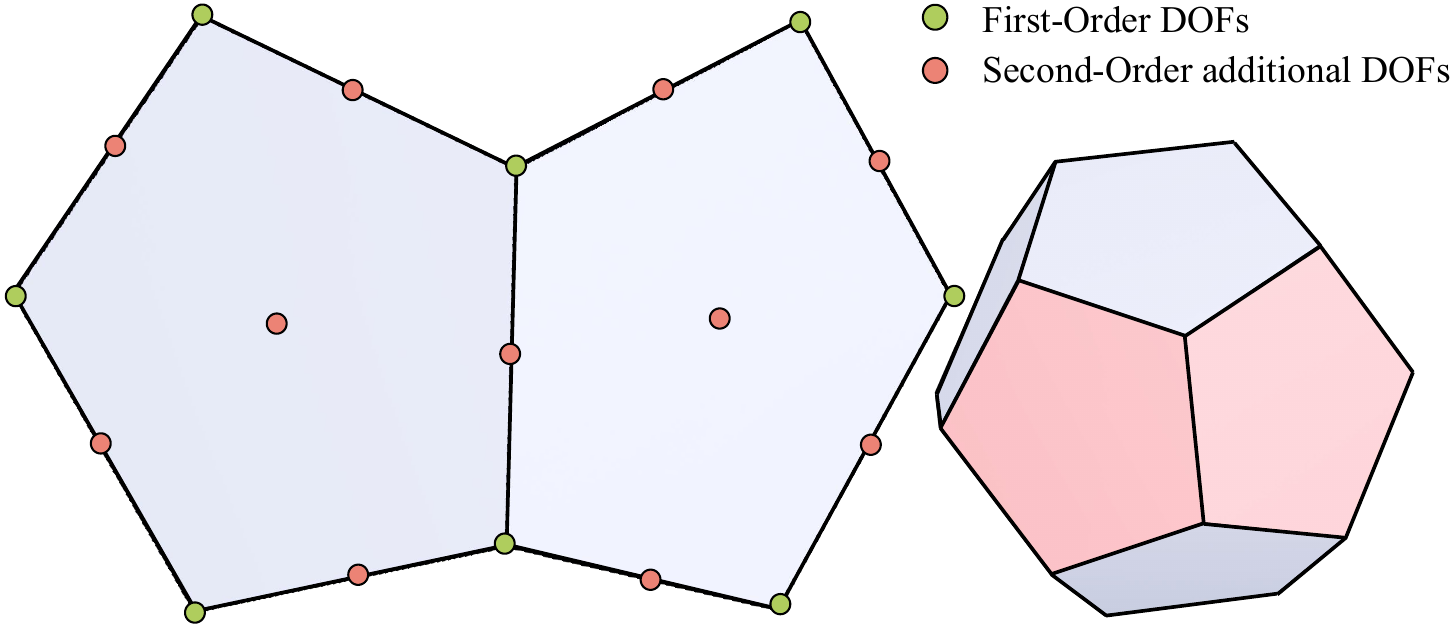}
    \caption{In the linear setting, VEM places degrees of freedom at grid vertices; in the
second-order formulation, additional degrees of freedom are
introduced on edges, faces, and within cells.
}
    \label{fig:dofs}
\end{figure}

\noindent
The resulting linear system involves only polynomial quantities on the
left-hand side and can therefore be evaluated exactly. 
To render the right-hand
side computable without  constructing $v_h$, the virtual element method
defines the DOFs of $v_h \in V_k(E)$ as follows:
\begin{enumerate}
    \item \emph{Vertex values}: the values of $v_h$ at the vertices of the
    polyhedral element $E$.
    \item \emph{Edge values}: on each edge $e \subset \partial E$, the values of
    $v_h$ at the $k-1$ internal points of the $(k+1)$-point Gauss--Lobatto quadrature.
    \item \emph{Face moments}: for each face $f \subset \partial E$, the moments
    of $v_h$ over $f$ up to order $k-2$,
    \[
    \frac{1}{|f|}\int_f v_h\, m_{\alpha}^f\,\mathrm{d}s,
    \qquad \alpha = 1,\dots,N_{k-2}^f.
    \]
    \item \emph{Cell moments}: the moments of $v_h$ over $E$ up to order $k-2$,
    \[
    \frac{1}{|E|}\int_E v_h\, m_{\alpha}\,\mathrm{d}x,
    \qquad \alpha = 1,\dots,N_{k-2}.
    \]
\end{enumerate}
Here, \( |f| \) and \( |E| \) denote the area of \( f \) and the volume of \( E \), respectively, and \( m_{\alpha}^f \) denotes the two-dimensional counterpart of \( m_{\alpha} \). Moreover, $N_{k-2}$ and
$N_{k-2}^f$ denote the numbers of basis polynomials of degree at most $k-2$ in the
three-dimensional and two-dimensional settings, respectively.  An intuitive illustration of the DOF placement is shown in Fig.~\ref{fig:dofs}. 
With the above definition of DOFs, the right-hand side terms can either be directly queried from the DOFs of \( v_h \) or computed using the DOFs via integration by parts. Consequently, the element-wise linear system corresponding to Eq.~\eqref{eq:VirtualProj} can be established and solved to obtain the projection operator \( \Pi_k^{\nabla} \), which maps virtual functions onto the polynomial space.

Using the projector \( \Pi_k^{\nabla} \), the basis function \( \varphi_i \) is decomposed as
\begin{equation}
\varphi_i = \Pi_k^{\nabla} \varphi_i + (I - \Pi_k^{\nabla}) \varphi_i.
\end{equation}
The element stiffness matrix can then be written as
\begin{equation}
K^E_{ij}
=
\int_E \nabla \Pi^{\nabla}_{k}\varphi_i \cdot \nabla \Pi^{\nabla}_{k}\varphi_j \,\mathrm{d}x
+
\int_E \nabla (I-\Pi^{\nabla}_{k})\varphi_i \cdot \nabla (I-\Pi^{\nabla}_{k})\varphi_j \,\mathrm{d}x .
\end{equation}
\noindent
where the cross terms vanish by the definition of $\Pi^{\nabla}_{k}$.
The first term ensures consistency and, since it involves only integrals with known polynomial functions, can be computed exactly. The second term provides stability and, following~\cite{vemtheory}, is approximated using the DOFs:
\begin{equation}
\sum_{i=1}^{N_{\mathrm{dof}}}
D_i\!\left((I-\Pi^{\nabla}_{k})\varphi_i\right)\,
D_i\!\left((I-\Pi^{\nabla}_{k})\varphi_j\right).
\end{equation}
Once $K^E$ is computed, the global stiffness matrix $K$ can be assembled.

\subsection{Particle-to-Grid Transfer}
\label{sec:p2g}
In our formulation, the particle-to-grid (P2G) step builds the DOFs of the first-order velocity field and assembles the right-hand-side vector $\vec r$, with entries
\begin{equation}
    \label{eq:right_hand_side}
\vec r_i=\frac{1}{\Delta t}\int_\Omega \mathbf u'_h \cdot \nabla \varphi_i\,\mathrm{d}x.
\end{equation}
As discussed in Section~\ref{sec:vemfluid:projection}, for a discrete scalar field, the first-order DOFs correspond to its values at vertices. Therefore, we adopt a PIC-style transfer to reconstruct velocity DOFs from particles.

The coefficient of the i-th degree of freedom of the discrete scalar field
\( f_h \) is computed via normalized radial kernel averaging:
\begin{equation}
\vec{f}_i = \frac{\sum_{p} f_p\, w_{ip}}{\sum_{p} w_{ip}},
\end{equation}
where \(f_p\) denotes the value of the scalar field carried by particle \( p \). For the weights \( w_{ip} \), we use a radial kernel function given by
\begin{equation}
w_{ip}
=
k\!\left(\|\mathbf{x}_p-\mathbf{x}_i\|^2 / R^2\right),
\qquad
k(q)=\max \bigl(0,(1-q)^3\bigr),
\end{equation}
where $\mathbf x_i$ denotes the position of the vertex associated with the $i$-th degree of freedom, and the kernel radius $R$ is set to the edge length of the Cartesian grid embedding the boundary. After P2G, the DOF vector $\vec{\mathbf{u}}$ of the velocity field $\mathbf{u}_h$ is obtained, from which the intermediate velocity $\mathbf{u}'_h$ is computed by applying external forces,
\begin{equation}
\mathbf{u}'_h = \mathbf{u}_h + \mathbf{f}_h\,\Delta t .
\end{equation}
Since both $\mathbf{u}_h$ and $\mathbf{f}_h$ lie in the same linear space $V_1$, the force contribution is added directly at the DOF vector level, with $\mathbf{f}_h$ computed using the Boussinesq buoyancy model.

% as
% \begin{equation}
% \mathbf{f} = (-\alpha\,\rho + \beta\,t)\,(0,1,0)^{\mathsf T}.
% \end{equation}
% \noindent

Given the intermediate velocity DOF vector $\vec{\mathbf{u}}'$, the right-hand-side term is evaluated via polynomial projection. Specifically, replacing $\mathbf{u}_h'$ with its projected polynomial $\Pi^{\nabla}_1 \mathbf{u}_h'$ in Eq.~\eqref{eq:right_hand_side} yields:
\begin{equation}
\frac{1}{\Delta t}\int_{\Omega} \mathbf{u}_h'\cdot\nabla v_i\,\mathrm{d}x
\;\approx\;
\frac{1}{\Delta t}\sum_{E\subset\Omega}
\int_{E} \Pi^{\nabla}_1 \mathbf{u}_h'\cdot\nabla v_i\,\mathrm{d}x.
\end{equation}
\noindent
Then, by the definition of the projection $\Pi^{\nabla}_k$, we have
\begin{equation}
\frac{1}{\Delta t}\int_{E} \Pi^{\nabla}_1 \mathbf{u}_h'\cdot\nabla v_i\,\mathrm{d}x
=
\frac{1}{\Delta t}\int_{E} \Pi^{\nabla}_1 \mathbf{u}_h'\cdot \Pi^{\nabla}_2 (\nabla v_i)\,\mathrm{d}x,
\end{equation}
which reduces the evaluation to a polynomial integral over each element and makes the right-hand-side vector $\vec r$ directly computable.

\subsection{Grid-to-Particle Transfer}
\label{sec:g2p}
Leveraging polynomial projection, we generalize the FLIP formulation to the conforming VEM scheme, resulting in improved preservation of fine-scale smoke details. Within each element, the projected pressure field admits a quadratic polynomial
expansion,
\begin{equation}
\Pi^{\nabla}_2 p_h^2
=
\sum_{\alpha=1}^{N_2} s^{p}_{\alpha}\, m_{\alpha}.
\end{equation}
Similarly, the projected intermediate velocity field is represented using
first-order polynomials as
\begin{equation}
\Pi^{\nabla}_1 \mathbf{u}_h^{\prime\,1}
=
\sum_{\alpha=1}^{N_1} \mathbf{s}^{\,\mathbf{u'}}_{\alpha}\, m_{\alpha}.
\end{equation}
A divergence-free velocity field is obtained by updating the polynomial
representation of the velocity,
\begin{equation}
\Pi^{\nabla}_1 \mathbf{u}^{\mathrm{new}}_h
=
\Pi^{\nabla}_1 \mathbf{u}_h^{\prime\,1}
-
\nabla \Pi^{\nabla}_2 p_h^2 .
\end{equation}
Since $\nabla \Pi^{\nabla}_2 p_h^2$ is itself a first-order polynomial, its
coefficients can be directly combined with those of
$\Pi^{\nabla}_1 \mathbf{u}_h^{\prime\,1}$.
After obtaining the updated velocity field in the polynomial space, we compute the velocity increment as:
\begin{equation}
\Pi^{\nabla}_1 \Delta \mathbf{u}_h
=
\Pi^{\nabla}_1 \mathbf{u}^{\mathrm{new}}_h
-
\Pi^{\nabla}_1 \mathbf{u}_h .
\end{equation}
The resulting velocity increment is transferred to particles using the standard FLIP scheme~\cite{pic}, with particle velocities updated as:
\begin{equation}
\mathbf{u}_p^{\mathrm{new}}
=
\alpha_{\mathrm{FLIP}}\,\mathbf{u}_p^{\mathrm{old}}
+
\Pi^{\nabla}_1 \Delta \mathbf{u}_h(\mathbf{x}_p)
+
(1-\alpha_{\mathrm{FLIP}})\,\Pi^{\nabla}_1 \mathbf{u}_h^{\mathrm{new}}(\mathbf{x}_p),
\end{equation}
where $\mathbf{u}_p^{\mathrm{old}}$ and $\mathbf{u}_p^{\mathrm{new}}$ denote the
particle velocity before and after the update, respectively, $\mathbf{x}_p$
denotes the particle position, and $\alpha_{\mathrm{FLIP}}\in[0,1]$ controls the
balance between FLIP and PIC updates. Following the velocity update, particle advection is performed using a
second-order Runge--Kutta scheme. 
During advection, an AABB tree is employed for collision detection, and mirror
reflection is applied to project particles back into the fluid domain.

\section{Galerkin Geometric Multigrid for VEM}
\label{sec:multigrid}
Designing Galerkin multigrid methods for VEM Poisson problems on general
polyhedral grids is particularly challenging, as Galerkin projection on
general polyhedral grids often induces significant connectivity proliferation among DOFs, leading to dense coarse-level system
matrices~\cite{Tiantian}. Here, connectivity refers to the
presence of nonzero entries in the system matrix coupling pairs of DOFs. 
This issue is exacerbated in second-order VEM on cut-cell grids for two main
reasons. First, second-order VEM exhibits dense local connectivity: vertex, edge, face, and
volume DOFs within each cell are all strongly coupled, thereby amplifying the
propagation of connectivity compared to first-order schemes. Second, cut-cell grids are
highly irregular, with some cells containing a large number of vertices, edges,
and faces, leading to substantially larger local stiffness matrices and making
the diffusion effect even more pronounced. 
To prevent connectivity proliferation from degrading multigrid efficiency, we
introduce a \emph{VEM-specific, diffusion-free prolongation operator} that
prevents cross-cell DOF connectivity, thereby avoiding dense
coarse-level system matrices.

\subsection{Multigrid Overview}

Multigrid methods are theoretically optimal and practically efficient  solvers for large sparse linear systems. Their effectiveness stems from a divide-and-conquer strategy that combines two complementary processes: relaxation and coarse-grid correction. Relaxation applies a smoother to eliminate high-frequency errors on the fine grid. Under the standard assumption that low-frequency errors on a fine grid appear as relatively higher-frequency errors on a coarser grid, coarse-grid correction transfers these errors to a coarser level via \emph{restriction}, applies smoothing there, and then propagates the correction back to the fine grid through \emph{prolongation}. 

Our method belongs to the class of geometric multigrid methods with Galerkin
coarse-grid approximation~\cite{MGTT}.
Unlike purely geometry-based multigrid schemes that re-discretize the governing
equations on coarser grids, both prolongation and restriction
operators in our method are constructed through the Galerkin
framework.
Consequently, the efficiency of the solver critically depends on the
construction of the grid hierarchy and the design of inter-grid transfer
operators. Galerkin multigrid constructs the coarse-level system matrix $A_c$ from the
fine-level matrix $A_f$ as
\begin{equation}
     \label{multigrid_construct}
A_c = R A_f P,
\end{equation}
where $R$ and $P$ denote the restriction and prolongation operators,
respectively.
To ensure variational consistency, Galerkin multigrid typically sets \( R = P^{\mathsf{T}} \), making the design of the prolongation operator crucial for the solver's efficiency. In this setting, the prolongation operator must
accurately interpolate smooth functions to achieve rapid convergence, while the prolongation matrix \(P\) should remain sparse to prevent densification
of coarse-level matrices.

\subsection{Naive VEM Prolongation Operator}
Assuming a hierarchy of polyhedral grids $\{\mathcal{P}^\ell\}_{\ell=1}^{L}$,
ordered from coarse to fine, with corresponding VEM spaces
$\{V^\ell\}_{\ell=1}^{L}$, prolongation maps a function
$v_h^{\ell-1}\in V^{\ell-1}$ on the coarse level to a function
$v_h^{\ell}\in V^{\ell}$ on the fine level.
We first attempt to generalize the definition of the prolongation operator used for first-order 2D VEM in ~\cite{vemgmg} to the second-order 3D VEM, resulting in the \emph{naive VEM prolongation operator}. This operator reconstructs the fine-level function from coarse-level degrees of freedom by directly transferring shared vertex DOFs and evaluating the remaining fine-level DOFs from the projected coarse-level function, i.e.,
\begin{equation}
    \label{eq:standard_prolongation}
    v_h^{\ell}
    :=
    \sum_{i\in\mathcal{N}(V^{\ell-1})}
    D_i\!\left(v_h^{\ell-1}\right)\,\varphi_i^{\ell}
    \;+\;
    \sum_{i\in\mathcal{N}(V^{\ell}\setminus V^{\ell-1})}
    D_i\!\left(\Pi^{\nabla}_2 v_h^{\ell-1}\right)\,\varphi_i^{\ell},
\end{equation}
where $\mathcal{N}(V^{\ell})$ denotes the index set of DOFs in $V^{\ell}$.
This prolongation requires computing the face- and volume-moments of a fine-level polyhedral cell, which, in turn, necessitates a \emph{nested} polyhedral grid hierarchy to ensure their efficient and consistent evaluation. Here, nested indicates that
each fine-level polyhedral cell is fully contained within a single
coarse-level cell.

Through the experiments, we found that the naive VEM prolongation operator induces a diffusion effect, leading to extremely dense coarse-level system matrices and poor efficiency when used either as a standalone solver or as a preconditioner. This diffusion effect is illustrated experimentally in
Fig.~\ref{fig:diffusion-free}, and a theoretical analysis of the phenomenon is
provided in Appendix~\ref{appendix:diffusion_analyze}.

\begin{figure}[t!]
    \centering
    \includegraphics[width=\linewidth]{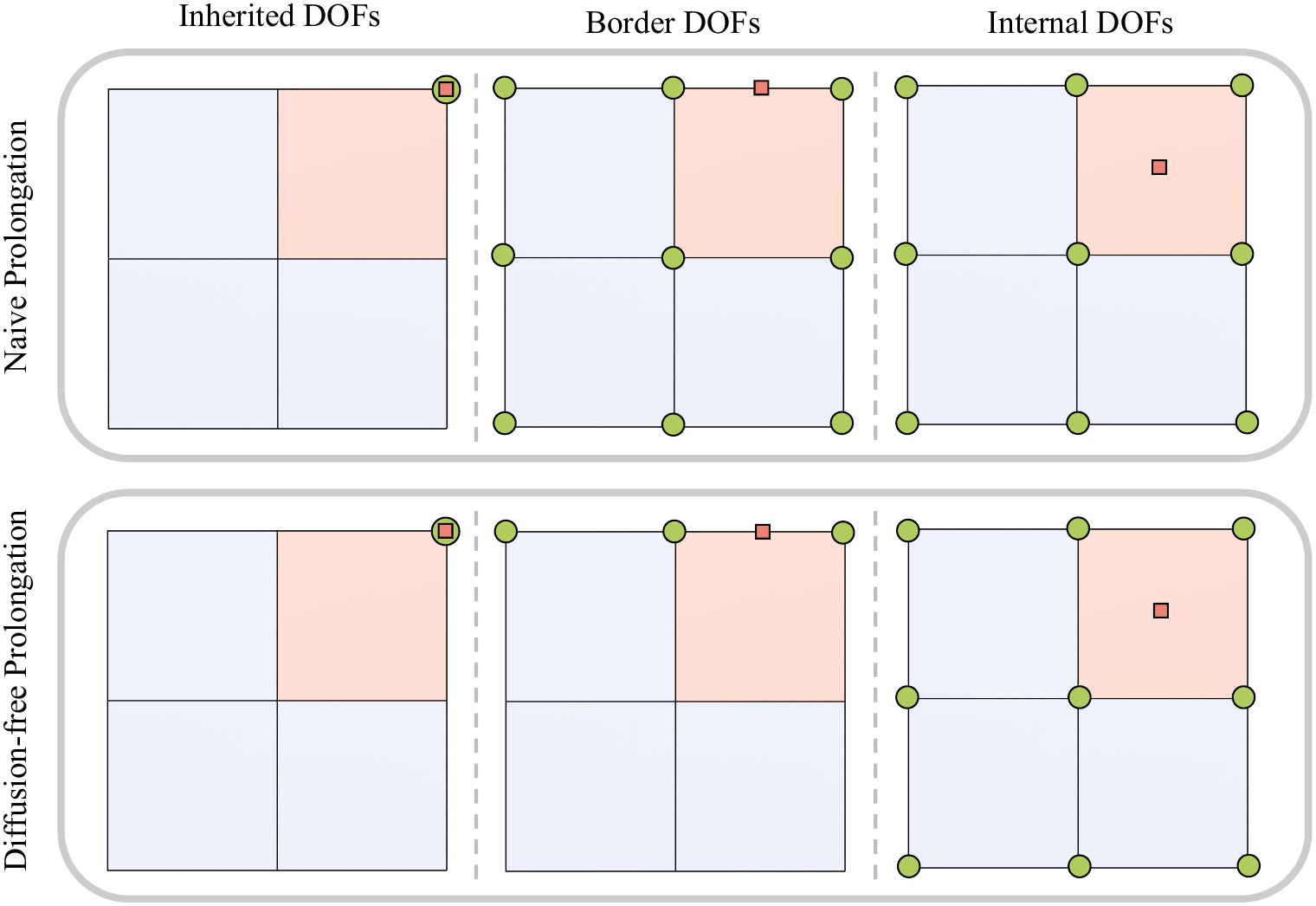}
    \caption{The large square represents a coarse-level cell, with green circles indicating coarse-level DOFs. The orange square denotes a fine-level cell, with orange markers indicating fine-level DOFs. To prolong error corrections from the coarse level to the fine level, coarse-level DOFs are used as interpolation points; the figure illustrates which coarse-level DOFs are involved in the prolongation of different categories of fine-level DOFs. Compared to the naive prolongation operator, our diffusion-free prolongation modifies the treatment of boundary DOFs (middle), effectively preventing the spurious spread of connectivity induced by boundary DOFs.}
    \label{fig:df_pro}
\end{figure}

\subsection{Diffusion-Free VEM Prolongation Operator}

To prevent connectivity between DOFs across cells at coarse levels, we propose a \emph{diffusion-free VEM prolongation operator},
defined as
\begin{equation}
\begin{aligned}
v_h^{\ell}
:={}&
\sum_{i\in\mathcal{N}(V^{\ell-1})}
D_i\!\left(v_h^{\ell-1}\right)\,\varphi_i^{\ell}
\\
&+
\sum_{i\in\mathcal{N}_e(V^{\ell}\setminus V{\ell-1})}
D_i\!\left(\Pi^{\nabla}_{2,e} v_h^{\ell-1}\right)\,\varphi_i^{\ell}
\\
&+
\sum_{i\in\mathcal{N}_f(V^{\ell}\setminus V{\ell-1})}
D_i\!\left(\Pi^{\nabla}_{2,f} v_h^{\ell-1}\right)\,\varphi_i^{\ell}
\\
&+
\sum_{i\in\mathcal{N}_v(V^{\ell}\setminus V{\ell-1})}
D_i\!\left(\Pi^{\nabla}_{2} v_h^{\ell-1}\right)\,\varphi_i^{\ell}.
\end{aligned}
\end{equation}
\noindent
Here, newly introduced fine-level DOFs are partitioned according to their
physical meaning.
Specifically, $\mathcal{N}_e$, $\mathcal{N}_f$, and $\mathcal{N}_v$ denote the
sets of edge, face, and volume DOFs, respectively.
These sets form a disjoint decomposition of the newly introduced fine-level DOFs:
\[
\mathcal{N}_e^{\ell}\cup\mathcal{N}_f^{\ell}\cup\mathcal{N}_v^{\ell}
=
\mathcal{N}(V^{\ell})\setminus\mathcal{N}(V^{\ell-1}),
\qquad
\mathcal{N}_e^{\ell}\cap\mathcal{N}_f^{\ell}\cap\mathcal{N}_v^{\ell}=\emptyset .
\]
The operators $\Pi^{\nabla}_{2,e}$ and $\Pi^{\nabla}_{2,f}$ denote quadratic
polynomial projections constructed via one- and two-dimensional VEM
formulations on coarse-level edges and faces, respectively.  These projections depend only on the DOFs associated with the corresponding
edges or faces, rather than on all DOFs of the polyhedral element, thereby ensuring a diffusion-free
prolongation. Under the proposed prolongation operator, the resulting coarse-level
matrix satisfies $A_c(i,j)\neq 0$ if and only if DOFs $i$ and $j$ belong to the
same polyhedral element at the coarse level.
The one- and two-dimensional VEM formulations are obtained by straightforward
dimensional reduction of the three-dimensional formulation described in
Section~\ref{sec:vemfluid:projection}.  

An intuitive comparison between the naive prolongation and the proposed
diffusion-free prolongation is shown in Fig.~\ref{fig:df_pro}. With the
prolongation operator defined, the multigrid hierarchy matrices are constructed
according to Eq.~\ref{multigrid_construct}. For high-frequency error reduction,
we employ a Chebyshev smoother with default parameters, implemented
in~\cite{AMGCL}. Compared to Jacobi smoothing, it is better suited to VEM
formulations and, unlike Gauss--Seidel smoothers, is straightforward to
parallelize on GPUs.

\section{Multilevel Body-Fitted Grid Construction}
\label{sec:cut-cell}
\begin{figure*}[t!]
    \centering
    \includegraphics[width=\linewidth]{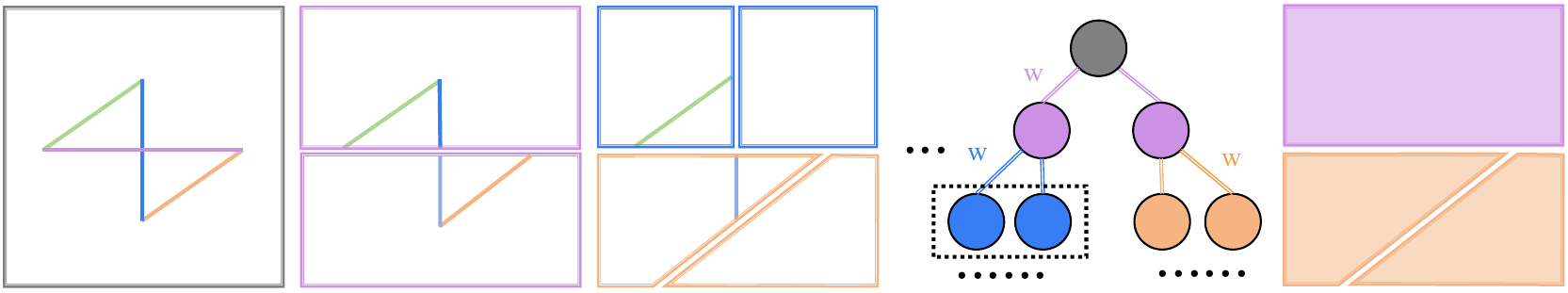}
    \caption{Illustration of a Binary Space Cut-Cell procedure and the corresponding multilevel grid construction. Starting from a regular gray cell, planar boundaries contained within the cell iteratively partition it through half-space cuts, resulting in body-fitted grids. This process is recorded as a binary tree, where the importance of each partition is quantified by an importance weight~$w$. Based on these weights, a greedy strategy removes cuts that are least beneficial for boundary embedding, yielding a coarsened grid. In this example, removing the cut highlighted by the dashed box produces the grid shown on the right.}
    \label{fig:cut_cell}
\end{figure*}

To support the diffusion-free geometric multigrid method described above,
a multilevel hierarchy of nested, body-fitted polyhedral grids is required.
This section describes the construction of body-fitted polyhedral grids and
their associated multilevel hierarchy. Specifically,
Section~\ref{sec:cutcell} presents the generation of a body-fitted grid from a
given boundary mesh, and Section~\ref{sec:multilevelgrid} introduces the
construction of the nested multilevel grid hierarchy. A illustrative example is provided in Fig.~\ref{fig:cut_cell} to facilitate understanding of the algorithm.

\subsection{Binary Space Cut-Cell}
\label{sec:cutcell}
Inspired by~\cite{CoLOD,LODTree}, we propose a simple and robust binary space
partitioning-based cut-cell strategy for constructing body-fitted polyhedral grids. Starting
from a background Cartesian grid, each boundary triangle is first associated
with all grid cells it intersects. This association is determined by uniformly
sampling points on each triangle and assigning the triangle to any grid cell
containing at least one sample.

For each grid cell, the associated triangles are ordered according to the number
of intersections between their supporting planes and other triangles, with those
exhibiting fewer intersections prioritized for half-space partitioning. This
ordering reduces the number of cut cells and, consequently, the size of the
resulting VEM stiffness matrix. Following this order, half-space partitioning is applied
independently within each grid cell: the supporting plane of the selected
triangle subdivides the current polyhedral cell into sub-cells, and the
remaining triangles are propagated to the corresponding sub-cells based on
their spatial relationship to the cutting plane. This procedure continues until all
relevant boundary triangles have been processed, yielding a body-fitted polyhedral
grid. During body-fitted grid construction, partitions that produce elements with volumes below $10^{-12}$ are discarded to
improve numerical robustness.

Each resulting polyhedron is subsequently classified as interior or
\setlength{\intextsep}{2pt}
\setlength{\columnsep}{5pt}
\begin{wrapfigure}{r}{0.15\textwidth}
	\includegraphics[
          width=0.15\textwidth,
          trim=1 0 0 0,
          clip
          ]{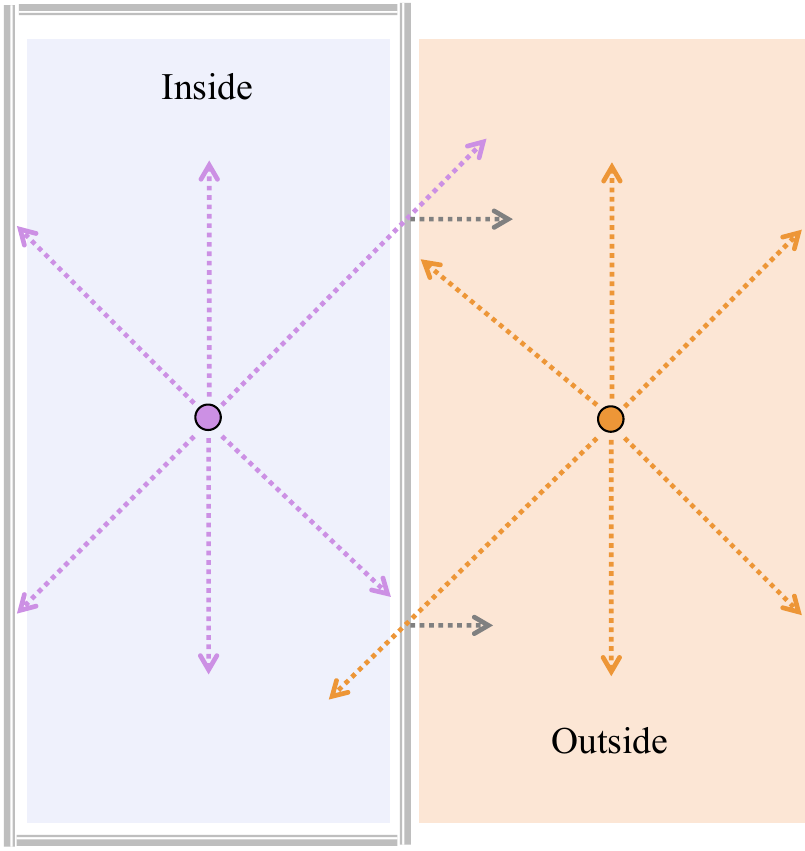}
\end{wrapfigure}
exterior with respect to the boundary mesh using ray casting. Rays are emitted from the centroid of
polyhedron and intersected with the boundary mesh; a polyhedron is
classified as interior if more than half of the rays intersect boundary
triangles whose normals form an angle smaller than $90^\circ$ with the ray
direction. An two-dimensional illustrative ray casting example is shown in right. Based
on this classification, only exterior polyhedra are retained for VEM
discretization, thereby naturally embedding boundary information into the grid
construction. Ray casting is accelerated using an AABB tree, with all intersection queries
handled by CGAL.

\subsection{Multilevel Grid Construction}
\label{sec:multilevelgrid}
As discussed in Section~\ref{sec:multigrid}, our multigrid method requires a
nested grid hierarchy capable of representing fine-level low-frequency errors on the coarser
grids.
This implies that coarse grids should remain as well aligned with the boundary
geometry as possible, so that smooth errors near boundaries can be effectively
represented through well-posed placement of coarse-level DOFs.

To satisfy these requirements, we propose a \emph{nested, boundary-aware} grid
hierarchy construction algorithm. Building on the cut-cell
procedure described in Section~\ref{sec:cutcell}, each initial regular cell is
treated as the root of a binary tree that records the sequence of half-space
partitions applied during grid construction. 
To enable boundary-aware hierarchy construction, each partition is assigned an
importance weight~$w$ that quantifies the contribution of the corresponding cut
to boundary embedding. Since a cut may introduce a polygonal face that is only partially aligned with the boundary geometry, the importance weight~\(w\) is used to quantify the boundary-covered portion of that face. It is defined as the area of the boundary-aligned portion divided by the total area of the boundary mesh.
These weights then guide a grid coarsening process, yielding a boundary-aware grid hierarchy. 

Coarsening is performed for $L-1$ iterations, yielding coarse-level polyhedral
grids $\{\mathcal{P}^{\ell}\}_{\ell=1}^{L-1}$. The transition from
$\mathcal{P}^{\ell}$ to $\mathcal{P}^{\ell-1}$ proceeds in two stages. First, regular cubic cells that do not contain cuts are aggregated following standard geometric multigrid practice, effectively halving the grid resolution.
Second, for irregular grid cells introduced by cuts, a greedy strategy is applied: removable partitions with the smallest importance weights are iteratively eliminated until the accumulated removed weight exceeds \(1/L\), where \(L\) denotes the total number of levels.
A partition is considered removable if its two child cells correspond to leaf
nodes; removing it merges the two cells into a single cell and
coarsens the grid locally.

After coarsening, the grid $\mathcal{P}^{\ell-1}$ is extracted from the remaining
cells. A cell is included in \(\mathcal{P}^{\ell-1}\) if it intersects the region enclosed by the boundary mesh.
 Since each coarse cell can be expressed as the union of a
set of finest-level cells, this test is evaluated efficiently by checking
whether any of the corresponding finest cells is classified as interior. The
interior/exterior classification at the finest level is obtained via ray
casting, as described in Section~\ref{sec:cutcell}.

By construction, the resulting grid hierarchy is nested: for any
$c^{\ell}\in\mathcal{P}^{\ell}$, there exists a cell
$c^{\ell-1}\in\mathcal{P}^{\ell-1}$ such that
$c^{\ell}\subset c^{\ell-1}$, and each coarse cell contains at least one finer
cell. This property prevents rank-deficient restriction operators and avoids
the introduction of spurious nullspaces in coarse-level system matrices.

\begin{figure*}[]
    \centering
    \includegraphics[width=\linewidth]{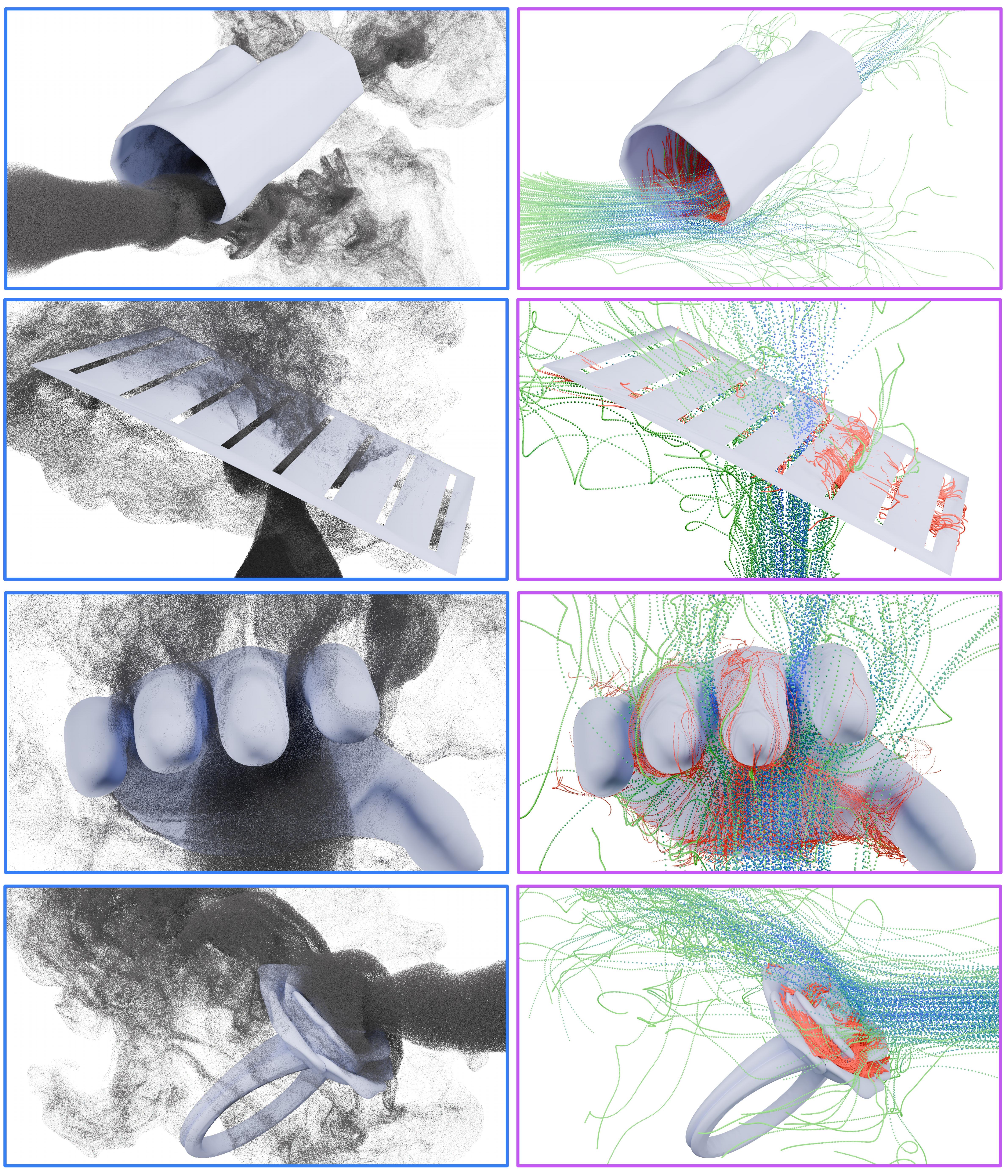}
    \caption{Smoke simulation results (left) and smoke particle trajectory visualizations
colored by velocity magnitude (right), where blue indicates higher speeds.
Near-surface flow behavior is visualized using red particles (right). From top to bottom:
\emph{Short}, \emph{Plane}, \emph{Hand}, and \emph{Rose}.
}
    \label{fig:results1}
\end{figure*}

\section{Results}
\label{sec:results}
\setlength{\tabcolsep}{4.5pt}  % 默认 6pt
\begin{table*}[]
\caption{Statistics of the 14 scenes used in our experiments. \emph{Faces} denotes the
number of triangles in the input boundary mesh. \emph{Grid Construction} reports
the time required to build the corresponding body-fitted grid. \emph{NNZs}
indicates the number of nonzero entries in the resulting stiffness matrix.
\emph{CD} denotes the bidirectional Chamfer distance between the extracted cut
mesh and the input surface, measuring boundary approximation error. 
\emph{Precompute} reports the additional time required to construct the
geometric multigrid hierarchy.}
\label{tab:scene}
\centering
\rowcolors{1}{white}{gray!15}
\begin{tabular}{cccccccccc}
\hline
Scene        & Features                        & Faces    & Grid Construction & NNZs & CD       & Precompute & P2G & Project & Advection  \\ \hline
Bunny        & Manifold, Closed                & $69451$  & $93.6$s                & $8.9e^8$     & $4.4e^{-5}$ & 42.3s          & 9.9s               &41.6s          & 8.9s                       \\
Dragon       & Self-Isects                     & $100000$ & $201.3$s                 & $9.4e^8$     & $4.7e^{-5}$ &  49.1s        & 11.6s              & 41.2s         &  9.3s                       \\
Hand         & Manifold, Closed                & $10856$  & $13.7$s                 & $6.5e^8$     & $2.8e^{-5}$ &38.0s         &  6.3s             &   20.6s      &    8.2s                     \\
Rose         & Self-Isects, Thin features      & $35905$  & $105.4$s                 & $7.5e^8$     & $3.2e^{-5}$ &31.9s          &  7.1s             &  25.4s        &   8.4s                      \\
Skull        & Self-Isects, Thin features      & $24808$  & $43.6$s                 & $7.1e^8$     & $3.6e^{-5}$ & 41.6s         &  8.7s             &   23.1s         &    8.0s                     \\
Ball         & Manifold, Closed                & $16380$  & $42.3$s                 & $6.4e^8$     & $2.2e^{-5}$ & 37.9s         &  7.9s             &   33.5s       &     7.8s                    \\
Hilbert      & High genus                      & $31060$  & $36.9$s                 & $8.1e^8$     & $1.9e^{-5}$ &  38.7s        &  10.0s             &  37.6s        &    8.6s                     \\
Short        & Thin features                   & $9168$   & $15.0$s                 & $6.7e^8$      & $1.8e^{-5}$ & 29.4s         &  5.3s             &    34.2s      &      7.1s                   \\
Lucy         & Manifold, Closed                & $99970$  & $128.5$s                & $9.0e^8$     & $3.9e^{-5}$ & 43.2s         &  9.7s             & 39.2s         &      8.8s                   \\
Hair Brush   & Self-Isects, Thin features      & $17442$  & $27.3$s                  &$6.8e^8$      & $2.3e^{-5}$ &  33.5s          & 11.2s             &  31.8s        &   9.3s                      \\
Conical Pipe & Non-Manifold, Thin features     & $15762$  & $2.9$s                 & $6.7e^8$     & $8.5e^{-6}$ & 31.0s          &   6.8s           &      25.3s      &                7.5s        \\
Plane             & Thin features              &$172$     & $3.8$s                 &$6.1e^8$      &$3.5e^{-6}$          &27.1s          &6.1s              &23.4s         & 7.7s                       \\
Panel             & Thin features              &$32571$   & $5.7$s                 &$6.3e^8$      &$4.1e^{-6}$          & 35.8s         & 7.5s              &   28.3s       &  8.1s                      \\
Building         & Non-Manifold, Self-Isects, Open &$1908013$          & $235.7s$                 & $6.2e^8$     & $2.9e^{-3}$         & 29.3s          & 7.7s              & 17.3s         & 6.3s                        \\ \hline
\end{tabular}
\end{table*}
\setlength{\tabcolsep}{8pt}  % 默认 6pt

\begin{table}[]
\caption{Comparison with existing cut-cell--based fluid simulators in terms of runtime for
the pressure projection stage. We implement a first-order pressure solver (denoted as Ours$^*$) for
comparison with~\cite{batty16}.\emph{Grid} indicates the resolution of the Cartesian background grid used for
embedding the boundary mesh.}
\label{tab:boundary_conforming}
\centering
\rowcolors{1}{white}{gray!15}
\begin{tabular}{cccc}
\hline
Method                      & Pressure Order & Grid    & Project \\ \hline
\cite{batty16} & 1              & $128^3$ & 47.2 s  \\
\cite{vempic}  & 2              & $50^3$  & 457.6 s \\
Ours$^*$                    & 1              & $128^3$ & 0.4 s    \\
Ours                        & 2              & $50^3$  & 3.7 s    \\ \hline
\end{tabular}
\end{table}

We evaluate FastVEM on 14 scenes spanning a broad range of complex boundary
configurations; the corresponding simulation settings and timings are summarized in Table~\ref{tab:scene}. In all cases, the boundary surface meshes are embedded into a $128^3$ Cartesian
background grid to construct body-fitted grids, and a five-level hierarchy is
built for the multigrid solver. FLIP particles are initialized according to the background grid resolution, with eight particles placed per grid cell. Particles located outside the simulation domain $\Omega$ (i.e., inside the boundary mesh) are marked as ghost particles to reduce P2G bias, and are excluded from both G2P transfer and advection. \re{In our setup, one time step corresponds to one frame. During advection, we employ substepping with a CFL number of 1 and grid spacing $\Delta x = 1/N$, where $N$ denotes the background grid resolution. The Advection column in Table~\ref{tab:scene} reports the total advection cost per time step, including all substeps.}

For each scene, we provide two complementary visualizations: a conventional
smoke rendering and a particle trajectory--based velocity visualization. The latter encodes particle speed using a colormap, with \textcolor{blue}{blue} indicating \textcolor{blue}{higher velocities}, and additionally highlights particles traveling near solid boundaries using \textcolor{red}{red-colored} particles to emphasize \textcolor{red}{near-boundary flow behavior}. \re{Green/blue pathlines trace 500–1000 randomly selected smoke particles, while red pathlines trace 100–300 particles that remain near solid boundaries for more than 50 consecutive frames.}
All experiments were conducted on a machine equipped with an Intel i9-13900K CPU, 64 GB of RAM, and an RTX 3090 GPU. Exact intersection computations required for cut-cell are handled using CGAL~\cite{cgal}.

\begin{figure*}[]
    \centering
    \includegraphics[width=\linewidth]{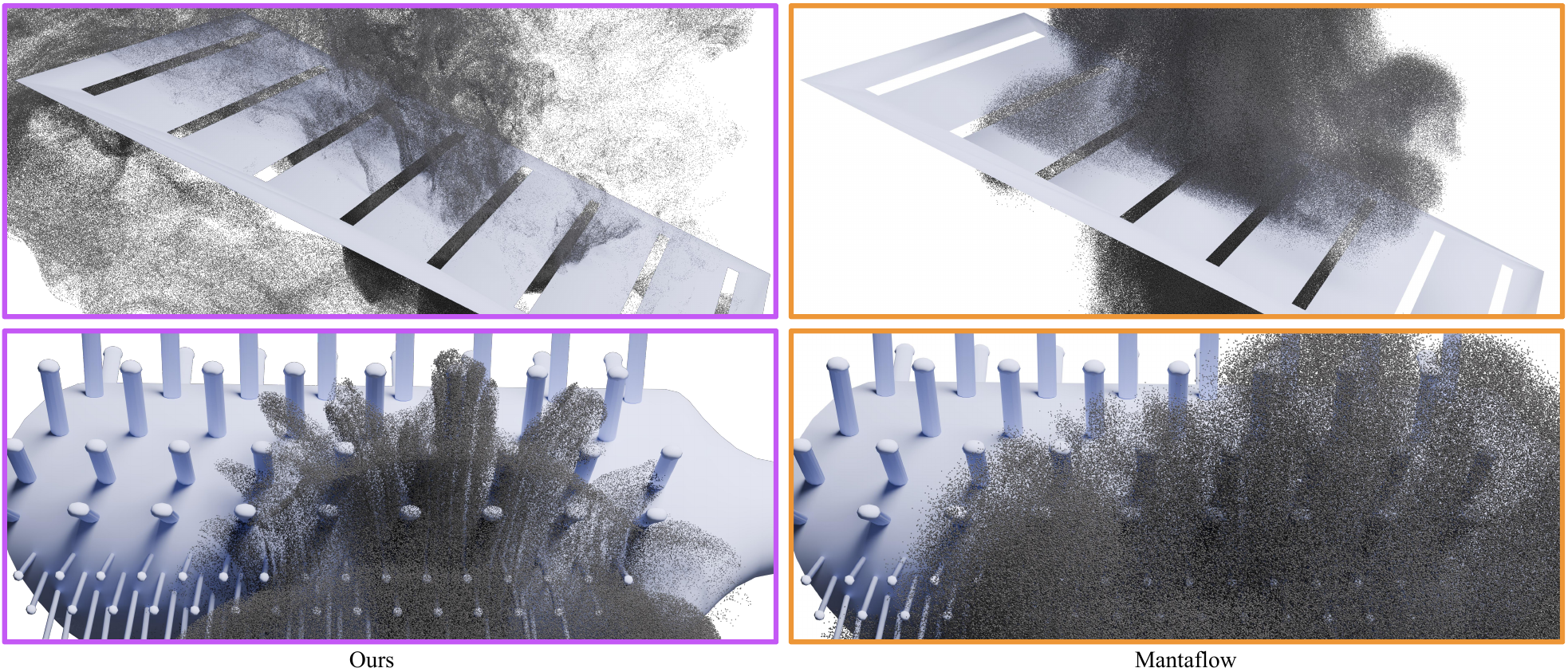}
    \caption{Comparison with Mantaflow. FastVEM more accurately resolves boundary interactions
involving thin structures and narrow gaps (top), as well as sub-grid curved
features (bottom), whereas Mantaflow fails to resolve these fine-scale geometric details.}
    \label{fig:Mantaflow}
\end{figure*}

\begin{figure}[t!]
    \centering
    \includegraphics[width=\linewidth]{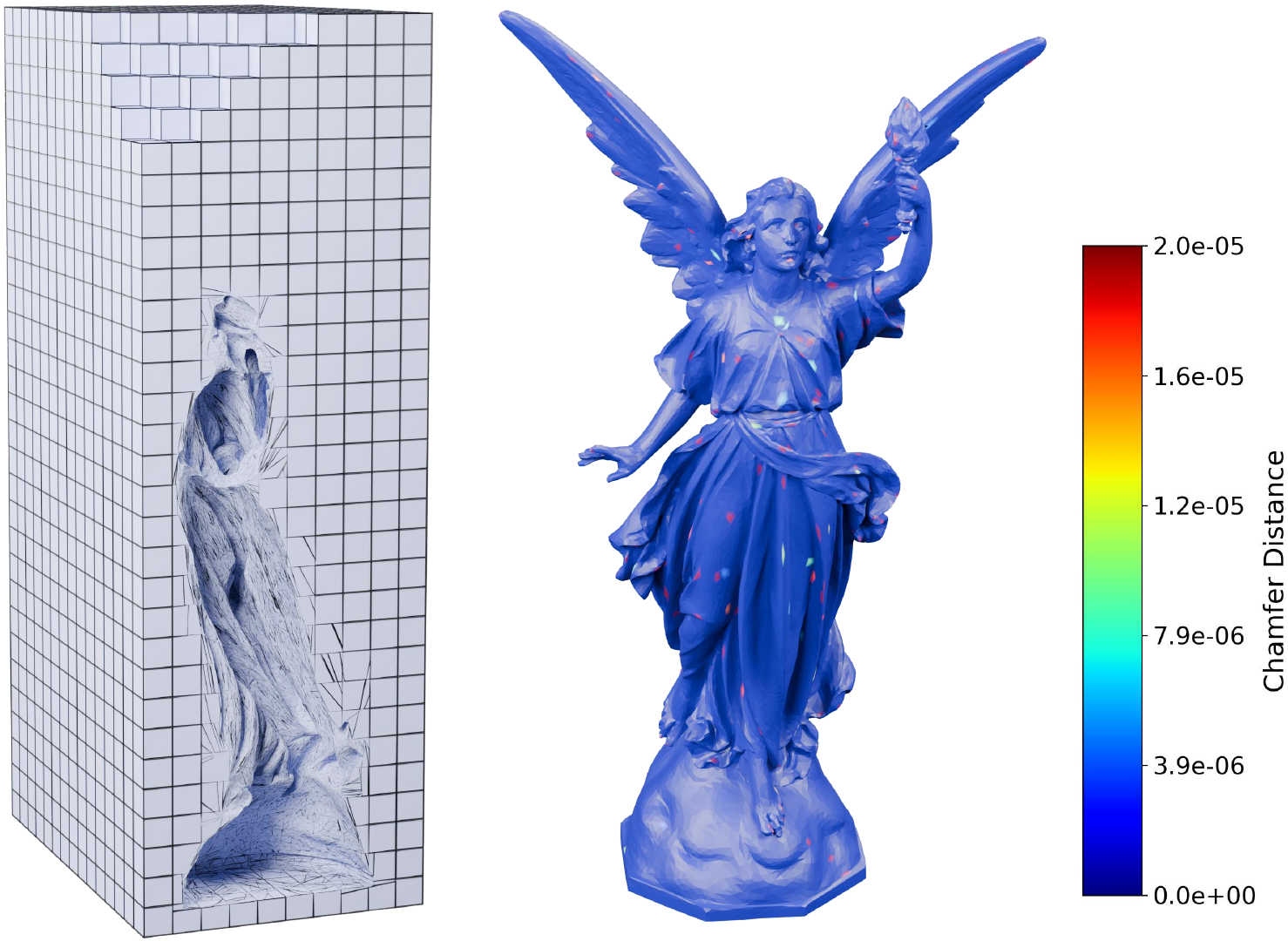}
    \caption{\re{Embedding of the Lucy model into a Cartesian grid using the proposed Binary Space Cut-Cell method.  The Chamfer distance between the embedded boundary and the original surface is visualized on the right, demonstrating the high geometric accuracy of the embedding.}}
    \label{fig:grid}
\end{figure}

\subsection{Complex Boundary Handling}

\begin{figure*}[]
    \centering
    \includegraphics[width=\linewidth]{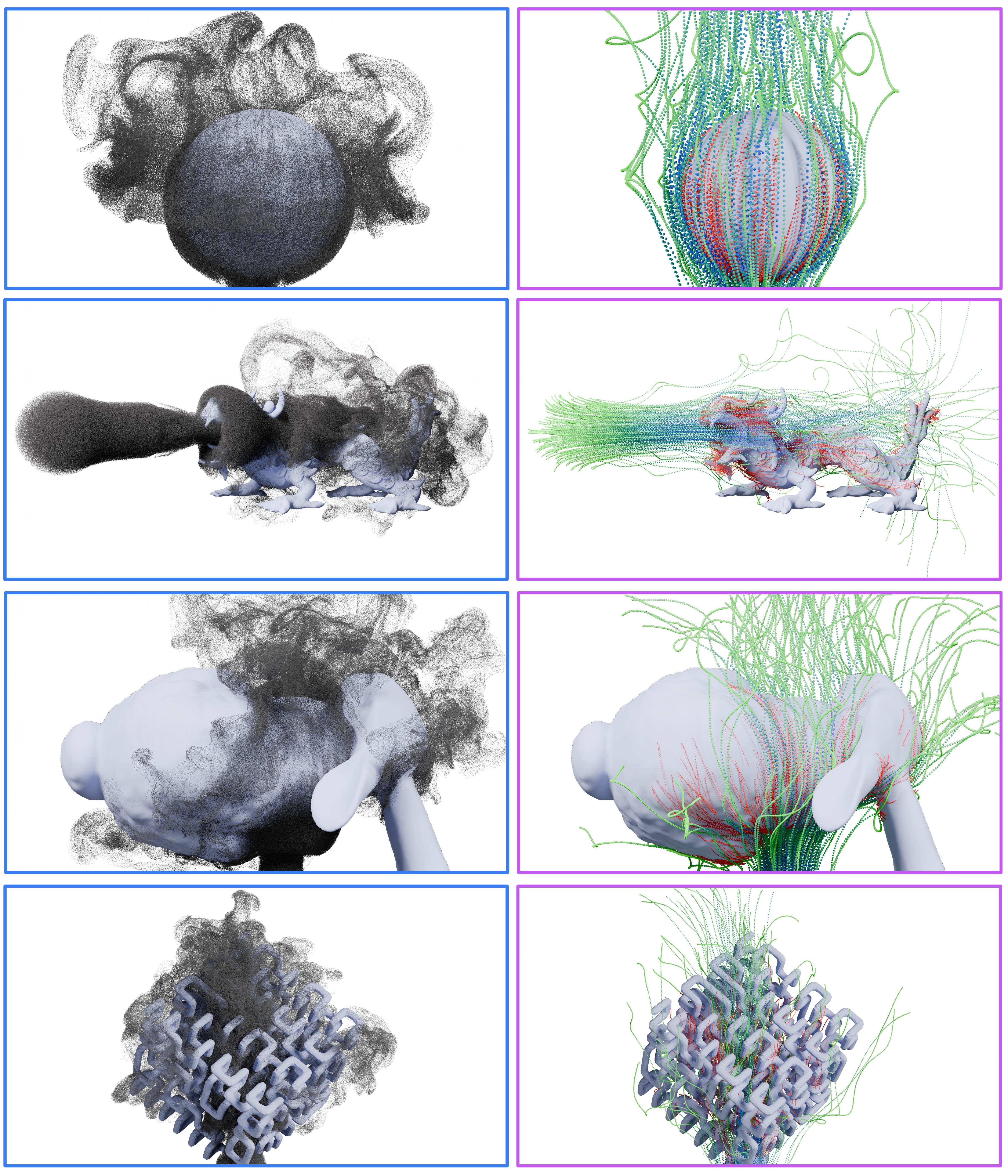}
    \caption{Smoke simulation results (left) and smoke particle trajectory visualizations
colored by velocity magnitude (right), where blue indicates higher speeds.
Near-surface flow behavior is visualized using red particles (right). From top to bottom:
\emph{Ball}, \emph{Dragon}, \emph{Bunny}, and \emph{Hilbert}.
}
    \label{fig:results2}
\end{figure*}

\begin{figure*}[]
    \centering
    \includegraphics[width=\linewidth]{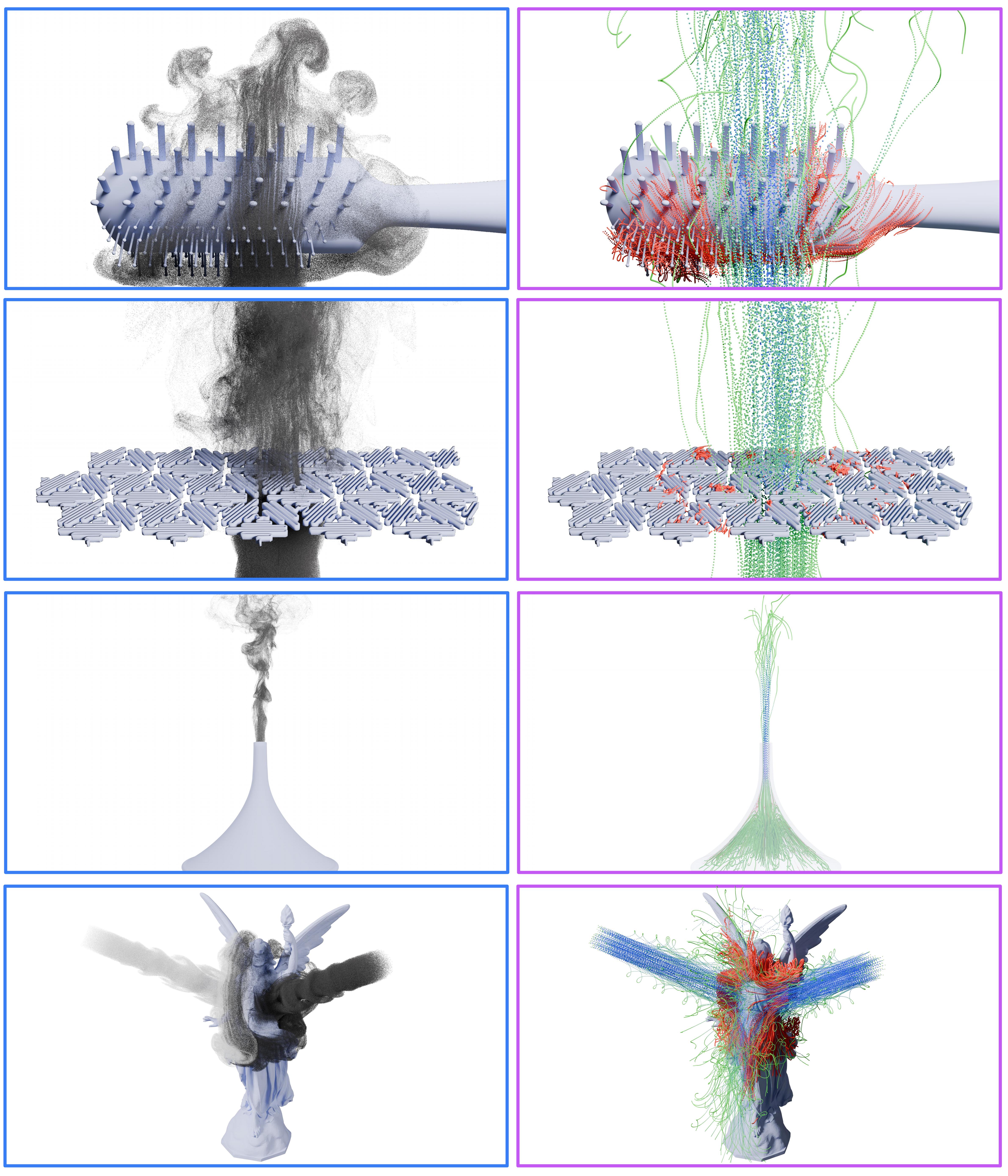}
    \caption{Smoke simulation results (left) and smoke particle trajectory visualizations
colored by velocity magnitude (right), where blue indicates higher speeds.
Near-surface flow behavior is visualized using red particles (right). From top to bottom:
\emph{Brush}, \emph{Panel}, \emph{Conical Pipe}, and \emph{Lucy}.
}
    \label{fig:results3}
\end{figure*}

FastVEM robustly resolves sub-grid boundary features that are challenging for
Cartesian grid methods, including thin structures
(Fig.~\ref{fig:results1}), narrow gaps
(Fig.~\ref{fig:teaser}; Fig.~\ref{fig:results3}), and highly curved surfaces
(Fig.~\ref{fig:results2}).
To more closely examine this advantage, we compare FastVEM against a Cartesian
grid--based fluid solver on the \emph{Brush} and \emph{Plane} scenes, both of which
contain thin structures and sub-grid geometric details. For this comparison, we
use Blender's built-in smoke simulator, Mantaflow~\cite{mantaflow}, configured
with a grid resolution of $256^3$ to provide a more challenging reference.

As shown in Fig.~\ref{fig:Mantaflow}, the smoke motion produced by Mantaflow
exhibits limited sensitivity to sub-grid boundary features. In the
\emph{Brush} scene, Mantaflow fails to capture the influence of thin,
rod-like structures on the flow, whereas FastVEM generates filamentary smoke
patterns that accurately reflect the flow perturbations induced by these
structures. The \emph{Plane} scene provides an even more illustrative comparison.
We set the plane thickness to $10^{-5}$ times the side length of the cubic
simulation domain, a regime that Mantaflow is essentially unable to handle: the smoke passes through the plane
with little apparent interaction, as if the boundary were absent. In contrast,
FastVEM produces boundary-conforming flow behavior, with the fluid being
deflected by the plane while portions of the flow pass through the narrow gaps.
For reference, the thickness of a real sheet of paper is on the order of
$0.1\,\mathrm{mm}$, indicating that FastVEM can resolve the influence of
paper-scale boundaries within simulation domains on the order of
$10\,\mathrm{m} \times 10\,\mathrm{m} \times 10\,\mathrm{m}$.

To further quantitatively evaluate the boundary-handling accuracy of FastVEM, we
measure the geometric error between the generated cut-cell mesh and the input
surface mesh, as reported in Table~\ref{tab:scene}. The error is quantified using
the bidirectional Chamfer Distance (CD), computed on meshes after normalized
scaling to remove global scale differences. Here, the cut-cell mesh refers to the
surface mesh extracted from the interior volumetric grid via ray casting,
following the procedure described in
Section~\ref{sec:cut-cell}. The results show that our cut-cell strategy achieves
errors on the order of $10^{-5}$ to $10^{-6}$, indicating that it can reliably
capture extremely fine sub-grid boundary features. \re{Fig.~\ref{fig:grid} further provides a visual illustration of the body-fitted grid and the boundary embedding error.}

% \subsection{Comparisons and Analysis}
\subsection{Computational Performance Comparison}

This subsection evaluates the performance of FastVEM through a series of
controlled comparisons. We first benchmark FastVEM against representative
boundary-conforming fluid solvers under matched simulation settings, where it
achieves speedups exceeding two orders of magnitude in the most time-consuming
pressure projection stage. We then analyze the sources of this acceleration via
an ablation-style study, comparing the proposed simulation-friendly
cut-cell strategy with Mandoline~\cite{mandoline} and quantitatively assessing
the performance gains provided by the multigrid solver.

\paragraph{\textbf{Comparison with boundary-conforming fluid solvers.}}
Table~\ref{tab:boundary_conforming} compares the computational efficiency of
FastVEM with closely related cut-cell--based boundary-conforming fluid solvers
under comparable experimental settings. We report the runtime of each method in
the pressure projection stage. All methods use the same boundary mesh
(\emph{Bunny}), and the reported results are taken directly from their published
papers. To ensure a fair comparison with the first-order pressure formulation of
Batty et~al.~\cite{batty16}, we implement a first-order VEM pressure solver within
our framework. As shown in the table, FastVEM demonstrates a substantial
efficiency advantage, achieving speedups of approximately two orders of
magnitude over both first-order and second-order boundary-conforming pressure
solvers.

\paragraph{\textbf{Comparison with Mandoline}}Compared to Mandoline, the proposed binary space partitioning--based cut-cell
strategy offers two key advantages. First, in our method, all generated grid
cells are guaranteed to be convex, avoiding non--star-shaped elements; this
improves grid quality and, in turn, leads to better conditioning of the
resulting VEM stiffness matrices. Second, the strategy prevents the
accumulation of boundary faces within individual grid cells, yielding a more
favorable local degree-of-freedom structure and, consequently, a sparser
global stiffness matrix.

Following~\cite{sigc}, we assess the quality of the body-fitted grids using a
VEM-specific element quality indicator. For each grid cell $E$, the
overall quality measure is defined as
\begin{equation}
    \label{eq:quality}
    \rho^{3\mathrm{D}}(E)
    =
    \frac{
    \sqrt{
    \rho^{3\mathrm{D}}_{1}(E)\,\rho^{3\mathrm{D}}_{2}(E)
    +
    \rho^{3\mathrm{D}}_{1}(E)\,\rho^{3\mathrm{D}}_{3}(E)
    }
    }{2},
\end{equation}
which combines three complementary factors to characterize the geometric
regularity and topological complexity of the generated grids. Detailed
definitions of the individual terms are provided in
Appendix~\ref{sec:quality_indicator}. Based on the element-wise quality measure, the overall quality of a body-fitted
polyhedral grid $\mathcal{P}$ is defined as the root-mean value of the quality
measures over all cut cells:
\begin{equation}
\rho(\mathcal{P})
=
\sqrt{
\frac{1}{\#\{E^c \in \mathcal{P}\}}
\sum_{E^c \in \mathcal{P}}
\rho^{3\mathrm{D}}(E^c)
}.
\end{equation}

\setlength{\tabcolsep}{2.2pt}  % 默认 6pt

\begin{table}[]
    \caption{Quantitative comparison of cut-cell grid quality and performance between
the proposed method and Mandoline~\cite{mandoline}.
We report grid quality (defined in Eq.~\ref{eq:quality}), global stiffness
matrix nonzeros (NNZs), pressure projection time, and cut-cell runtime.
For each entry, values are reported in the form \emph{Ours / Mandoline}.}

\label{tab:Mandoline}
\centering
\rowcolors{1}{white}{gray!15}
\begin{tabular}{ccccc}
\hline
Scene        & Quality & NNZs                & Project & Runtime       \\ \hline
Bunny        & 0.78 / 0.63        & $4.7e^8$ / $7.1e^8$         & 7.9s / 127.1s  & 84.2s / 4.6s  \\
Dargon       & 0.77 / 0.48        & $6.5e^8$ / $7.4e^8$         & 12.3s / 148.5s         & 187.1s / 5.4s \\
Hand         & 0.79 / 0.63        & $1.1e^8$ / $1.9e^8$          & 6.2s / 15.8s & 10.3s / 1.8s  \\
Rose         & 0.77 / 0.62        & $2.3e^8$ / $4.7e^8$         & 5.5s / failed  & 93.9s / 6.5s  \\
Skull        & 0.78 / 0.62        & $1.9e^8$ / $2.7e^8$         &   4.7s / 80.5s     & 38.9s / 6.7s  \\
Ball         & 0.78 / 0.64        & $1.0e^8$ / $2.0e^8$ & 1.7s / 221.8s  & 41.1s / 1.6s  \\
Hilbert      & 0.80 / 0.65        & $2.4e^8$ /  $3.8e^8$        & 7.2s / 103.2s  & 29.6s / 4.7s  \\
Short        & 0.79 / 0.74        & $9.4e^7$ / $1.5e^8$          & 3.1s / 5.9s & 11.8s / 2.1s  \\
Lucy         & 0.77 / 0.52        & $7.4e^8$ / $8.3e^8$            & 11.5s / 254.1s        & 178.9s / 5.1s \\
Hair Brush   & 0.79 / 0.52        & $1.3e^8$ / $2.3e^8$            & 3.5s / failed & 23.3s / 8.9s  \\
Conical Pipe & 0.79 / 0.74        & $1.2e^8$ / $2.4e^8$            & 8.8s / 21.2s & 16.4s / 3.0s  \\
Plane        & 0.85 / 0.75        & $4.6e^7$ / $1.2e^8$          & 8.3s / 9.1s & 1.6s / 0.4s   \\
Panel        & 0.81 / 0.61       & $8.3e^7$ / $3.8e^8$           & 6.7s / failed   & 3.4s / 2.9s   \\
Building     & 0.85 / -         &   $6.1e^7$ / -                   & 2.5s / -        & 218.2s / failed \\ \hline
\end{tabular}
\end{table}

\begin{figure}[t!]
    \centering
    \includegraphics[width=\linewidth]{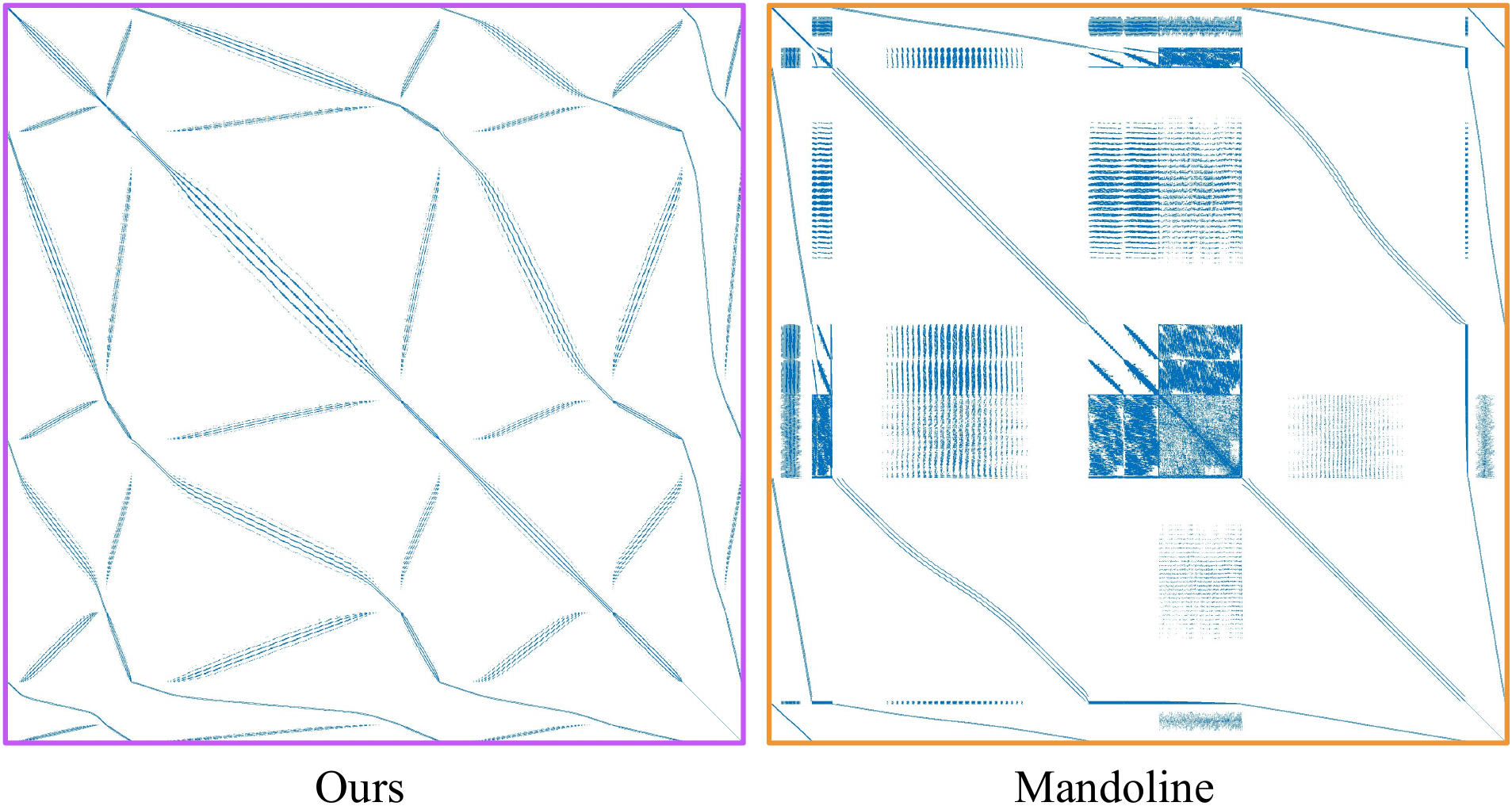}
    \caption{Compared to body-fitted grids constructed with Mandoline for VEM discretization,
our cut-cell strategy produces significantly sparser stiffness matrices and
avoids irregular dense substructures. This is a direct consequence of our
binary space partitioning strategy, which fundamentally prevents excessive
accumulation of boundary faces within individual cells.
}
    \label{fig:mandoline}
\end{figure}

\setlength{\tabcolsep}{7pt}  % 默认 6pt
\begin{table*}[]
        \caption{Performance comparison for solving Poisson systems arising from VEM
discretization. We compare DPCG and ICPCG using Eigen’s implementations, and
UA-AMGPCG using AMGCL. For both UA-AMGPCG and our GMGPCG, the Chebyshev smoother
in AMGCL is used with default parameters.
}
\label{tab:mg}
\centering
\rowcolors{1}{white}{gray!15}
\begin{tabular}{ccccccc}
\hline
Scene        & DPCG                 & ICPCG               & UA-AMGPCG             & Ours$^{nn}$        & Ours$^{np}$       & Ours             \\ \hline
Bunny        & 6975 iters / 3856.6s & 503 iters / 1532.0s & 140 iters / 111.0s & 102 iters / 301.2s & 13 iters / 107.5s & 16 iters / 41.6s \\
Dargon       & 7214 iters / 4725.3s & 602 iters / 2566.7  & 186 iters / 176.9s & 96 iters / 283.2s  & 14 iters / 153.0s & 15 iters / 41.2s \\
Hand         & 4632 iters / 2475.5s & 426 iters / 1231.2s & 75 iters / 58.8s   & 111iters / 348.9s  & 12 iters / 101.9s & 11 iters / 32.6s \\
Rose         & 3998 iters / 2474.1s & 325 iters / 1487.2s & 92 iters / 81.4s   & 105 iters / 273.8s & 9 iters / 75.6s   & 9 iters / 25.4s  \\
Skull        & 6325 iters / 4313.2s & 459 iters / 2629.1s & 168iters /  153.7s & 88 iters / 188.9s  & 11iters / 99.6s   & 13iters / 31.9s  \\
Ball         & 6322 iters / 3952.4s & 512 iters / 2240.4s & 85 iters / 60.6s   & 97 iters / 285.3s  & 12 iters / 108.4s & 11 iters / 33.5s \\
Hilbert      & 5326 iters / 3105.7s & 487 iters / 1831.3s & 110 iters / 90.1s  & 129 iters / 287.5s & 13 iters / 149.0s & 16 iters / 37.6s \\
Short        & 2316 iters / 1326.1s & 319 iters / 1021.6s & 167 iters / 104.6s & 71iters / 193.0s   & 12 iters / 72.4s  & 13 iters / 34.2s \\
Lucy         & 7013 iters / 4366.0s & 501 iters / 2593.2s & 196iters / 183.0s  & 130 iters / 88.7s  & 16iters / 201.7s  & 15iters / 39.2s  \\
Hair Brush   & 1721 iters / 1088.2s & 206 iters / 781.3s  & 131 iters / 96.3s  & 198 iters / 392.2s & 12 iters / 89.3s  & 14 iters / 31.8s \\
Conical Pipe & 3135 iters / 1750.8s & 399 iters / 1386.3s & 74 iters / 52.3s   & 67 iters / 237.8s  & 11 iters / 81.1s  & 11 iters / 25.3s \\
Plane        & 3845 iters / 2160.3s & 412 iters / 1620.2s & 217 iters / 129.4s & 17 iters / 41.0s   & 8 iters / 68.2s   & 8 iters / 23.4s  \\
Panel        & 2296 iters / 1421.9s & 267 iters / 1125.8s & 108 iters / 64.4s  & 39 iters / 78.7s   & 8 iters / 78.0s   & 10 iters / 28.3s \\
Building     & 2356 iters / 1517.5s                     & 231 iters / 916.9s                    & 52 iters /  42.6s                  & 68 iters / 153.2s                  & 7 iters / 61.1s                  &  7 iters / 17.3s                \\ \hline
\end{tabular}
\end{table*}

Table~\ref{tab:Mandoline} reports quantitative measures of body-fitted grid
quality, the number of nonzeros in the corresponding stiffness matrix, the
runtime of the pressure projection stage when simulating fluid flow on the
resulting grids, and the cut-cell runtime for both the proposed cut-cell
strategy and Mandoline. In this comparison, the boundary mesh is embedded into a
moderate $50^3$ background grid to ensure that the pressure projection cost on
Mandoline-generated grids remains tractable. Since Mandoline does not provide a
grid hierarchy, pressure projection on its grids is solved using UA-AMGPCG.

As shown in the first three columns of Table~\ref{tab:Mandoline}, FastVEM
consistently produces higher-quality body-fitted grids and sparser stiffness
matrices, which together lead to faster pressure projection.
Fig.~\ref{fig:mandoline} provides a direct visual comparison of the resulting
stiffness matrices, illustrating that Mandoline generates locally dense matrix
blocks, whereas our method yields a more favorable sparsity pattern.

Compared to Mandoline, our cut-cell strategy incurs a higher preprocessing cost
due to the use of exact geometric predicates for robust and accurate body-fitted
grid generation, as well as a currently non-parallel implementation. This
preprocessing, however, is performed only once and can be reused across
simulations with different physical parameters; consequently, its cost accounts
for only a small fraction of the overall simulation time.

\begin{figure}[]
    \centering
    \includegraphics[width=\linewidth]{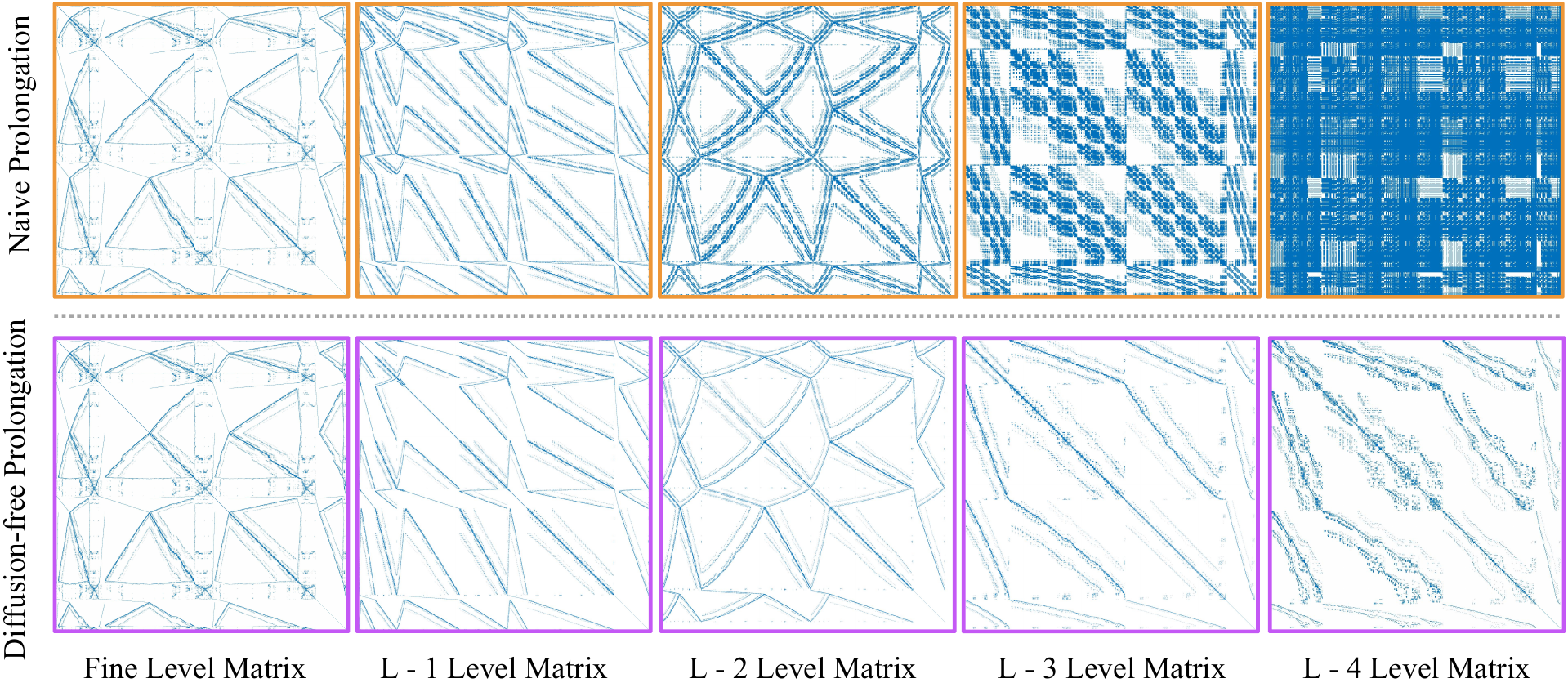}
    \caption{Compared to the naive prolongation operator obtained by directly extending a
two-dimensional first-order VEM multigrid method to three-dimensional
second-order settings, the proposed diffusion-free prolongation operator
effectively avoids coarse-level matrix densification.
}
    \label{fig:diffusion-free}
\end{figure}

\paragraph{\textbf{Multigrid Method Evaluation}} Table~\ref{tab:mg} summarizes the performance of our geometric multigrid
preconditioned conjugate gradient (GMGPCG) solver and several alternatives across
multiple scenes, including Diagonal Preconditioned Conjugate Gradient (DPCG),
Incomplete Cholesky Preconditioned Conjugate Gradient (ICPCG) using Eigen’s
implementation~\cite{eigen}, and Unsmoothed Aggregation Algebraic Multigrid Preconditioned Conjugate
Gradient (UA-AMGPCG) provided by AMGCL~\cite{AMGCL}. In all experiments, we construct five multigrid levels for our solver and set the
maximum number of levels in UA-AMGPCG to five as well. Under these matched settings,
FastVEM achieves an approximate $4\times$ speedup over the second-best method,
UA-AMGPCG. Notably, UA-AMGPCG exhibits unstable performance---particularly in scenes
with complex boundaries, where convergence degrades significantly---whereas the
proposed GMGPCG solver remains robust due to the boundary-aware grid hierarchy
construction.

We further perform ablation studies to assess the impact of two key components of
the proposed multigrid framework: the nested, boundary-aware LOD hierarchy
construction and the diffusion-free prolongation operator. As reported in
Table~\ref{tab:mg} (column \emph{Ours$^{nn}$}), replacing our hierarchy
construction with the strategy of~\cite{LODTree}, which neither enforces
nestedness nor explicitly accounts for boundary geometry, results in a substantial increase
in the number of iterations required for convergence and, consequently, higher
solution costs. Table~\ref{tab:mg} (column \emph{Ours$^{np}$}) further examines the effect of
naively extending the prolongation operator proposed in~\cite{vemgmg}, originally
designed for two-dimensional first-order VEM, to three-dimensional
second-order pressure solves, as defined in
Eq.~\eqref{eq:standard_prolongation}. This extension leads to coarse-level
matrices with significantly fill-in, as illustrated in
Fig.~\ref{fig:diffusion-free}. Although it requires slightly fewer conjugate gradient iterations, the
substantially higher per-iteration cost---caused by the reduced sparsity of
coarse-level matrices---ultimately renders this approach much less efficient
than the proposed diffusion-free prolongation--based multigrid method.

\begin{figure*}[]
    \centering
    \includegraphics[width=\linewidth]{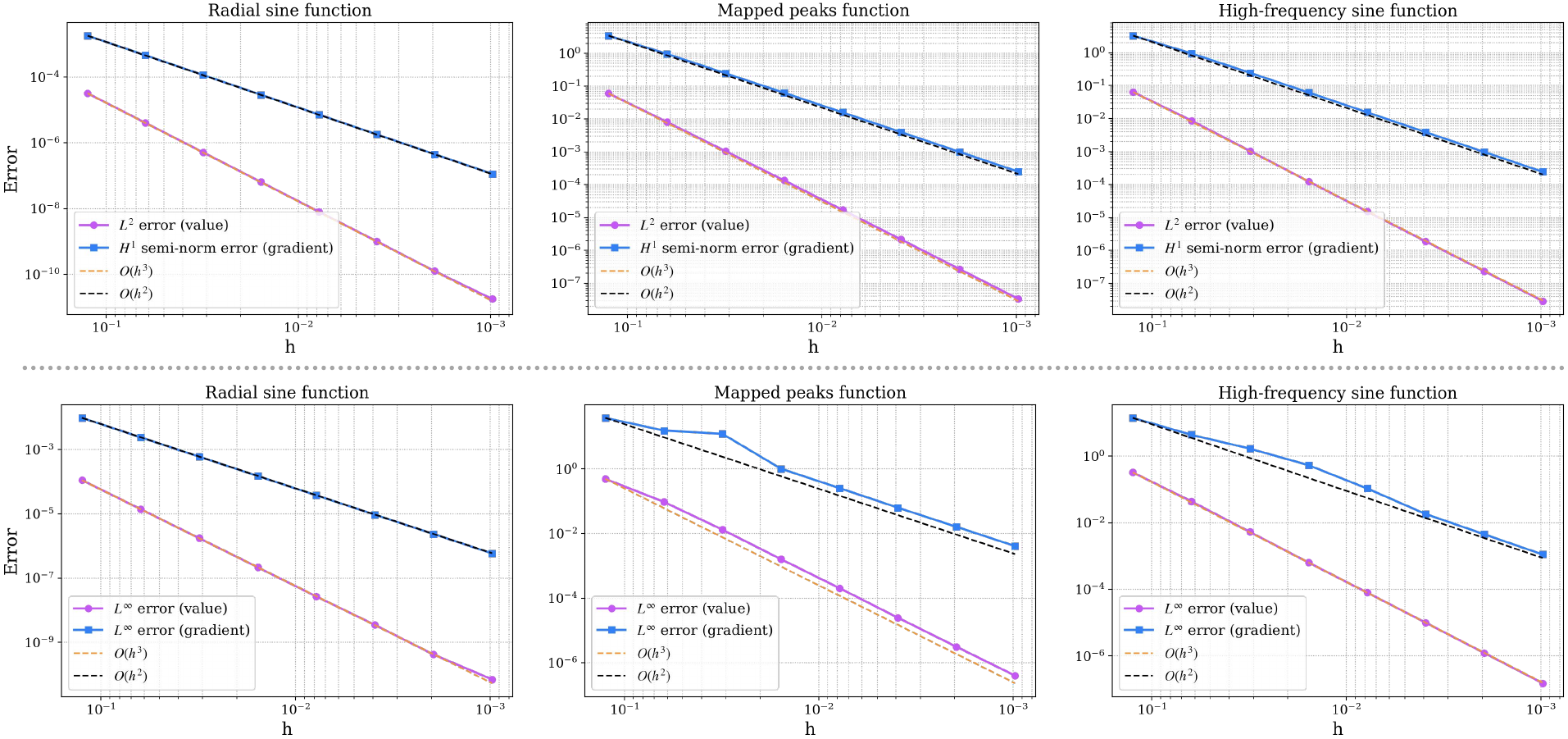}
    \caption{\re{Convergence study of second-order conforming VEM on three test functions (radial sine, mapped peaks, and high-frequency sine). 
Errors are plotted against the grid resolution parameter $h$ in log--log scale, where $h$ is defined as the inverse of the grid resolution (e.g., $h = 1/N$ for an $N \times N$ discretization). 
We report $L^2$ and $L^\infty$ errors for the solution, and the $H^1$ semi-norm and $L^\infty$ errors for its gradient. The slopes indicate approximately third-order convergence in the solution and second-order convergence in the gradient.
}}
    \label{fig:convergence}
\end{figure*}

\paragraph{\textbf{Convergence Tests.}}\re{ We evaluate the accuracy and convergence behavior of the second-order conforming VEM scheme by employing it as a Poisson solver for problems of the form
\begin{equation}
\begin{cases}
\nabla^2 f(x) = g(x), & x \in \mathcal{D}, \\
\frac{\partial f}{\partial n}(x) = h(x), & x \in \partial \mathcal{D},
\end{cases}
\end{equation}
where non-homogeneous Neumann boundary conditions are prescribed, consistent with pressure projection in incompressible flows.

Convergence tests are conducted on a $1 \times 1$ 2D domain using three representative functions. A circular obstacle centered at $(0.5, 0.5)$ with radius $0.2$, approximated by a piecewise-linear curve with 256 segments, is embedded into the Cartesian background grid. The test functions include a radial function $f(x,y)=r\sin r$ with $r=\sqrt{x^2+y^2}$; a mapped peaks function defined by evaluating the standard MATLAB peaks function under the transformation $X=6x-3,\,Y=6y-3$; and a high-frequency function $f(x,y)=\sin(kx)\sin(ky)$ with $k=5\pi$. As shown in Fig.~\ref{fig:convergence}, the second-order VEM exhibits approximately third-order convergence in the solution and second-order convergence in its gradient, in agreement with the theoretical convergence rates~\cite{vemtheory}.}

\begin{figure*}[]
    \centering
    \includegraphics[width=\linewidth]{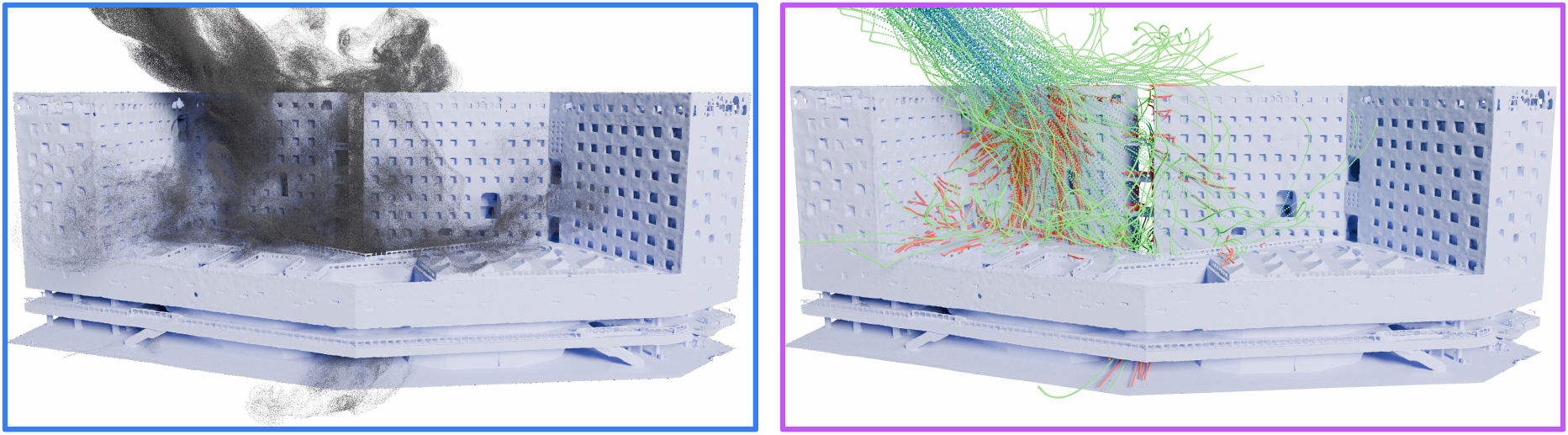}
    \caption{FastVEM supports fluid interaction with extremely complex and noisy models
reconstructed from drone scans.
}
    \label{fig:building}
\end{figure*}

\paragraph{\textbf{Further Applications.}} 
Our cut-cell strategy can be combined with planar detection techniques to
robustly handle extremely complex and noisy reconstructed meshes, for which
Mandoline fails to produce valid outputs. As shown in
Fig.~\ref{fig:building}, FastVEM successfully simulates smoke interacting with a
highly complex building model reconstructed from drone scans containing nearly
two million faces. More interestingly, as reported in
Table~\ref{tab:Mandoline}, the resulting body-fitted grids exhibit high element
quality, which in turn leads to reduced computational cost. Despite relying on
planar detection--based approximations, the representation accuracy remains on
the order of $10^{-3}$, as shown in Table~\ref{tab:scene}. The resulting
simulations accurately capture boundary-induced flow behavior: smoke is
deflected by complex architectural features, passes through narrow gaps, and
naturally emerges near the base of the structure.

\section{Disscussion and Future Works}
We present \textit{FastVEM}, a boundary-conforming fluid simulator on body-fitted
grids that \rre{speeds up the computationally dominant pressure projection stage by up to 100$\times$ over existing cut-cell--based fluid simulators.} FastVEM combines a generalized FLIP formulation for advection with a first-order
velocity and second-order pressure VEM discretization to robustly enforce
incompressibility and boundary conditions. Complementing this discretization,
we introduce a simple yet robust binary space partitioning--based cut-cell
strategy for constructing simulation-friendly body-fitted grids. This strategy
offers two key advantages: it guarantees convex grid cells, improving numerical
stability, and it avoids excessive clustering of boundary facets within
individual cells, thereby preventing overly dense local stiffness matrices.
To further accelerate pressure projection, we develop a Galerkin geometric multigrid
solver with a diffusion-free prolongation operator that prevents coarse-level
matrix densification, together with a nested, boundary-aware grid hierarchy that
improves convergence. Extensive experiments demonstrate that FastVEM robustly
handles a wide range of complex boundaries while maintaining practical
computational cost.

Our work is subject to several limitations. First, despite robust and efficient handling of complex boundary geometries,
FastVEM is currently limited to static boundaries. \re{Applying our framework to deforming boundaries or liquid simulations requires frequent grid updates to track evolving surfaces, incurring tens of seconds of overhead per timestep. This overhead stems from the lack of hierarchical body-fitted grid construction algorithms that preserve both mesh quality and efficiency, which remains an open problem. A highly parallel, floating-point--based binary space cut-cell strategy, instead of our current serial implementation relying on CGAL's exact intersection computations, could potentially mitigate this cost. Second, as shown in Table~\ref{tab:scene}, second-order VEM discretization produces up to $9.4\times10^8$ nonzeros at $128^3$, requiring 14~GB of GPU memory for the solver, which limits scalability to higher resolutions. Further exploration of more efficient stiffness matrix storage for regular-grid regions within the body-fitted grid may alleviate this limitation.} Third, while FastVEM can handle boundaries with open surfaces by implicitly sealing them during the ray-casting stage, this design choice limits its ability to support intentionally open surface features.
Nevertheless, interactions with objects typically modeled as open surfaces, such as cloth or planar sheets, can be simulated by representing them as thin shells with finite thickness, as illustrated in Fig.~\ref{fig:results1}. Extending FastVEM to directly support intentionally open surfaces could broaden the range of fluid--boundary interactions, but would require specialized partitioning of degrees of freedom near open boundaries and is therefore left for future investigation.

Looking ahead, FastVEM opens several promising directions for future research.
First, extending the FastVEM framework to support dynamic boundaries and, more
broadly, fluid--solid coupling could enable substantially richer visual effects.
Second, as demonstrated by our experiments, constructing simulation-friendly
body-fitted grids is far from trivial and leaves considerable room for further
exploration; for example, investigating how the ordering of space partitioning
in our cut-cell strategy affects grid quality could lead to improved
performance. \rre{Third, extending our VEM-specific multigrid design to other discretizations is a promising direction. VEM provides locally redundant DOFs with respect to the underlying polynomial space, enabling reduced global coupling in the prolongation operator while preserving interpolation quality, thereby supporting a diffusion-free construction while preserving V-cycle efficiency. Extending this design to other discretizations may require introducing auxiliary DOFs to balance coarse-level sparsity and V-cycle performance.
} Fourth, FastVEM adopts a Chebyshev smoother rather than a Jacobi smoother because,
for second-order VEM discretizations of the Poisson system, naive Jacobi
smoothers fail to converge for both our GMGPCG and UA-AMGPCG solvers. This
behavior likely arises because different diagonal entries encode distinct
physical meanings and therefore exhibit disparate scales and spectral
characteristics. Developing VEM-specific, highly parallel smoothers may thus be
key to further improving the efficiency of boundary-conforming fluid
simulators. Finally, many applications could benefit from explicitly embedding
boundaries into volumetric grids, including fluidic system design and
boundary-layer modeling, and we look forward to exploring such integrations in
future work.

% \textcolor{blue}{Looking ahead,xxx}
% \textcolor{blue}{Looking ahead, several promising directions remain for future exploration.
% As discussed in Section~\ref{sec:xxx}, an important next step is to extend FastVEM
% to support moving boundaries. Another compelling direction is the further
% development of VEM-specific cut-cell strategies for constructing higher-quality
% body-fitted grids, which are critical for improving the robustness and
% efficiency of VEM-based fluid simulators.
% In addition, body-fitted polyhedral grids open up new opportunities, including
% more natural boundary-layer modeling and improved accuracy for differentiable
% fluidic system design, making them promising avenues for future research.}

\begin{acks}
This work was supported by the Natural Science Foundation of China (62595772, 62272245), and the Fundamental Research Funds for the Central Universities (Nankai University, 63263248).
\end{acks}

\bibliographystyle{ACM-Reference-Format}
\bibliography{FastVEM_ref}

@STRING{SIGGRAPH = {Proc. SIGGRAPH}}

@STRING{TOG = {ACM Trans. on Graphics}}

@STRING{TVCG = {IEEE Trans. Visualization \& Computer Graphics}}

@STRING{ISPRS = {ISPRS J. Photogrammetry and Remote Sensing}}

@article{octree03,
author = {Losasso, Frank and Gibou, Fr\'{e}d\'{e}ric and Fedkiw, Ron},
title = {Simulating water and smoke with an octree data structure},
year = {2004},
issue_date = {August 2004},
publisher = {Association for Computing Machinery},
volume = {23},
number = {3},
journal = TOG,
month = aug,
pages = {457–462},
numpages = {6}
}

@article{powd_octree,
author = {Aanjaneya, Mridul and Gao, Ming and Liu, Haixiang and Batty, Christopher and Sifakis, Eftychios},
title = {Power diagrams and sparse paged grids for high resolution adaptive liquids},
year = {2017},
issue_date = {August 2017},
publisher = {Association for Computing Machinery},
volume = {36},
number = {4},
journal = TOG,
month = jul,
articleno = {140},
numpages = {12}
}

@article{octree20,
author = {Ando, Ryoichi and Batty, Christopher},
title = {A practical octree liquid simulator with adaptive surface resolution},
year = {2020},
volume = {39},
number = {4},
journal = TOG,
month = aug,
articleno = {32},
numpages = {17}
}

@article{whirp,
author = {Zhang, Xinxin and Li, Minchen and Bridson, Robert},
title = {Resolving fluid boundary layers with particle strength exchange and weak adaptivity},
year = {2016},
volume = {35},
number = {4},
journal = TOG,
month = jul,
articleno = {76},
numpages = {8}
}

@article{plbm,
author = {Liu, Mengyun and Liu, Xiaopei},
title = {A Parametric Kinetic Solver for Simulating Boundary-Dominated Turbulent Flow Phenomena},
year = {2023},
volume = {42},
number = {6},
journal = TOG,
month = dec,
articleno = {189},
numpages = {20},
}

@ARTICLE{overset,
  author={Xiao, Xiaoyu and Lin, Ding and Wu, Yiheng and Bai, Kai and Liu, Xiaopei},
  journal=TVCG, 
  title={Simulating Two-Phase Fluid-Rigid Interactions With an Overset-Grid Kinetic Solver}, 
  year={2025},
  volume={31},
  number={10},
  pages={8397-8412}
  }

@inproceedings{ghost,
  title={Using the particle level set method and a second order accurate pressure boundary condition for free surface flows},
  author={Enright, Doug and Nguyen, Duc and Gibou, Frederic and Fedkiw, Ron},
  booktitle={Fluids Engineering Division Summer Meeting},
  volume={36975},
  pages={337--342},
  year={2003}
}

@article{batty07,
  title={A fast variational framework for accurate solid-fluid coupling},
  author={Batty, Christopher and Bertails, Florence and Bridson, Robert},
  journal=TOG,
  volume={26},
  number={3},
  pages={100--es},
  year={2007},
  publisher={ACM New York, NY, USA}
}

@article{areafractions,
title = {An efficient fluid–solid coupling algorithm for single-phase flows},
journal = {Journal of Computational Physics},
volume = {228},
number = {23},
pages = {8807-8829},
year = {2009},
author = {Yen Ting Ng and Chohong Min and Frédéric Gibou},
}

@article{icutmg,
author = {Weber, Daniel and Mueller-Roemer, Johannes and Stork, André and Fellner, Dieter},
title = {A Cut-Cell Geometric Multigrid Poisson Solver for Fluid Simulation},
journal = {Computer Graphics Forum},
volume = {34},
number = {2},
pages = {481-491},
year = {2015}
}

@article{batty10,
author = {Batty, Christopher and Xenos, Stefan and Houston, Ben},
title = {Tetrahedral Embedded Boundary Methods for Accurate and Flexible Adaptive Fluids},
journal = {Computer Graphics Forum},
volume = {29},
number = {2},
pages = {695-704},
year = {2010}
}

@article{edwards14,
author = {Edwards, Essex and Bridson, Robert},
title = {Detailed water with coarse grids: combining surface meshes and adaptive discontinuous Galerkin},
year = {2014},
volume = {33},
number = {4},
journal = TOG,
month = jul,
articleno = {136},
numpages = {9}
}

@article{batty16,
author = {Azevedo, Vinicius C. and Batty, Christopher and Oliveira, Manuel M.},
title = {Preserving geometry and topology for fluid flows with thin obstacles and narrow gaps},
year = {2016},
volume = {35},
number = {4},
journal = TOG,
month = jul,
articleno = {97},
numpages = {12}
}

@inproceedings{batty17,
author = {Zarifi, Omar and Batty, Christopher},
title = {A positive-definite cut-cell method for strong two-way coupling between fluids and deformable bodies},
year = {2017},
booktitle = {Proceedings of the ACM SIGGRAPH / Eurographics Symposium on Computer Animation},
articleno = {7},
numpages = {11},
series = {SCA '17}
}

@article{lbm21,
author = {Lyu, Chaoyang and Li, Wei and Desbrun, Mathieu and Liu, Xiaopei},
title = {Fast and versatile fluid-solid coupling for turbulent flow simulation},
year = {2021},
volume = {40},
number = {6},
journal = TOG,
month = dec,
articleno = {201},
numpages = {18}
}

@article{vempic,
author = {Tao, Michael and Batty, Christopher and Ben-Chen, Mirela and Fiume, Eugene and Levin, David I. W.},
title = {VEMPIC: particle-in-polyhedron fluid simulation for intricate solid boundaries},
year = {2022},
volume = {41},
number = {4},
journal = TOG,
month = jul,
articleno = {115},
numpages = {22}
}

@article{mandoline,
author = {Tao, Michael and Batty, Christopher and Fiume, Eugene and Levin, David I. W.},
title = {Mandoline: robust cut-cell generation for arbitrary triangle meshes},
year = {2019},
volume = {38},
number = {6},
journal = TOG,
month = nov,
articleno = {179},
numpages = {17}
}

@inproceedings{McAdams,
author = {McAdams, A. and Sifakis, E. and Teran, J.},
title = {A parallel multigrid Poisson solver for fluids simulation on large grids},
year = {2010},
booktitle = {Proceedings of the 2010 ACM SIGGRAPH/Eurographics Symposium on Computer Animation},
pages = {65–74},
numpages = {10},
series = {SCA '10}
}

@ARTICLE{Chentanez2012,
  author={Chentanez, Nuttapong and Mueller-Fischer, Matthias},
  journal=TVCG, 
  title={A Multigrid Fluid Pressure Solver Handling Separating Solid Boundary Conditions}, 
  year={2012},
  volume={18},
  number={8},
  pages={1191-1201}
  }

@article{tallgrid,
author = {Chentanez, Nuttapong and M\"{u}ller, Matthias},
title = {Real-time Eulerian water simulation using a restricted tall cell grid},
year = {2011},
volume = {30},
number = {4},
journal = TOG,
month = jul,
articleno = {82},
numpages = {10},
}

@article{SPGrid,
author = {Setaluri, Rajsekhar and Aanjaneya, Mridul and Bauer, Sean and Sifakis, Eftychios},
title = {SPGrid: a sparse paged grid structure applied to adaptive smoke simulation},
year = {2014},
volume = {33},
number = {6},
journal = TOG,
month = nov,
articleno = {205},
numpages = {12}
}

@article{Dick2016,
  title={Solving the Fluid Pressure Poisson Equation Using Multigrid—Evaluation and Improvements},
  author={Christian Dick and Marcus Rogowsky and R{\"u}diger Westermann},
  journal=TVCG,
  year={2016},
  volume={22},
  pages={2480-2492}
}

@article{Aanjaneya,
author = {Aanjaneya, Mridul and Han, Chengguizi and Goldade, Ryan and Batty, Christopher},
title = {An Efficient Geometric Multigrid Solver for Viscous Liquids},
year = {2019},
volume = {2},
number = {2},
journal = {Proc. ACM Comput. Graph. Interact. Tech.},
month = jul,
articleno = {14},
numpages = {21}
}

@inproceedings{Zarifi,
author = {Zarifi, Omar},
title = {Sparse Smoke Simulations in Houdini},
year = {2020},
booktitle = {ACM SIGGRAPH 2020 Talks},
articleno = {3},
numpages = {2},
series = {SIGGRAPH '20}
}

@article{amgfluid,
author = {Shao, Han and Huang, Libo and Michels, Dominik L.},
title = {A fast unsmoothed aggregation algebraic multigrid framework for the large-scale simulation of incompressible flow},
year = {2022},
volume = {41},
number = {4},
journal = TOG,
month = jul,
articleno = {49},
numpages = {18},
}

@article{AMGCL,
title = {AMGCL —A C++ library for efficient solution of large sparse linear systems},
journal = {Software Impacts},
volume = {6},
pages = {100037},
year = {2020},
author = {Denis Demidov},
}

@book{MGTT,
author = {Briggs, William L. and Henson, Van Emden and McCormick, Steve F.},
title = {A multigrid tutorial (2nd ed.)},
year = {2000},
publisher = {Society for Industrial and Applied Mathematics}
}

@article{Tiantian,
author = {Xian, Zangyueyang and Tong, Xin and Liu, Tiantian},
title = {A scalable galerkin multigrid method for real-time simulation of deformable objects},
year = {2019},
volume = {38},
number = {6},
journal = TOG,
month = nov,
articleno = {162},
numpages = {13}
}

@article{LJM,
author = {Lu, Jia-Ming and Yuan, Tailing and Mo, Zhe-Han and Hu, Shi-Min},
title = {Fast Galerkin Multigrid Method for Unstructured Meshes},
year = {2025},
volume = {44},
number = {6},
journal = TOG,
month = dec,
articleno = {179},
numpages = {16}
}

@article{vempmg,
     author = {Antonietti, Paola F. and Mascotto, Lorenzo and Verani, Marco},
     title = {A multigrid algorithm for the p-version of the virtual element method},
     journal = {ESAIM: Mathematical Modelling and Numerical Analysis },
     pages = {337--364},
     year = {2018},
     volume = {52},
     number = {1}
}

@article{vemgmg,
author = {Antonietti, Paola F. and Berrone, Stefano and Busetto, Martina and Verani, Marco},
title = {Agglomeration-Based Geometric Multigrid Schemes for the Virtual Element Method},
journal = {SIAM Journal on Numerical Analysis},
volume = {61},
number = {1},
pages = {223-249},
year = {2023}
}

@article{fvm,
  title={Finite volume methods},
  author={Eymard, Robert and Gallou{\"e}t, Thierry and Herbin, Rapha{\`e}le},
  journal={Handbook of numerical analysis},
  volume={7},
  pages={713--1018},
  year={2000},
  publisher={Elsevier}
}

@article{HDG,
author = {Cockburn, Bernardo and Gopalakrishnan, Jayadeep and Lazarov, Raytcho},
title = {Unified Hybridization of Discontinuous Galerkin, Mixed, and Continuous Galerkin Methods for Second Order Elliptic Problems},
journal = {SIAM Journal on Numerical Analysis},
volume = {47},
number = {2},
pages = {1319-1365},
year = {2009}
}

@article{WG,
  title={A weak Galerkin finite element method for second-order elliptic problems},
  author={Wang, Junping and Ye, Xiu},
  journal={Journal of Computational and Applied Mathematics},
  volume={241},
  pages={103--115},
  year={2013},
  publisher={Elsevier}
}

@article{HHO,
title = {An Arbitrary-Order and Compact-Stencil Discretization of Diffusion on General Meshes Based on Local Reconstruction Operators},
author = {Daniele A. Di Pietro and Alexandre Ern and Simon Lemaire},
pages = {461--472},
volume = {14},
number = {4},
journal = {Computational Methods in Applied Mathematics},
year = {2014}
}

@article{PFM,
author = {Sukumar, N. and Tabarraei, A.},
title = {Conforming polygonal finite elements},
journal = {International Journal for Numerical Methods in Engineering},
volume = {61},
number = {12},
pages = {2045-2066},
year = {2004}
}

@article{MFD,
     author = {Brezzi, Franco and Buffa, Annalisa and Lipnikov, Konstantin},
     title = {Mimetic finite differences for elliptic problems},
     journal = {ESAIM: Mod\'elisation math\'ematique et analyse num\'erique},
     pages = {277--295},
     year = {2009},
     volume = {43},
     number = {2}
}

@inproceedings{sigc,
author = {Sorgente, Tommaso and Vicini, Fabio and Cabiddu, Daniela and Biasotti, Silvia and Spagnuolo, Michela and Manzini, Gianmarco and Berrone, Stefano},
title = {Mesh Quality Meets The Virtual Element Method},
booktitle = {SIGGRAPH Asia 2024 Courses},
articleno = {9},
numpages = {93},
year = {2024},
series = {SA Courses '24}
}

@article{vemtheory,
  title={Basic principles of virtual element methods},
  author={Beir{\~a}o da Veiga, Lourenco and Brezzi, Franco and Cangiani, Andrea and Manzini, Gianmarco and Marini, L Donatella and Russo, Alessandro},
  journal={Mathematical Models and Methods in Applied Sciences},
  volume={23},
  number={01},
  pages={199--214},
  year={2013}
}

@article{CoLOD,
author = {Zhang, Runze and Pan, Shanshan and Lv, Chenlei and Gong, Minglun and Huang, Hui},
title = {Architectural Co-LOD Generation},
year = {2024},
volume = {43},
number = {6},
journal = TOG,
month = nov,
articleno = {193},
numpages = {16}}

@article{LODTree,
title = {Building LOD representation for 3D urban scenes},
journal = {ISPRS Journal of Photogrammetry and Remote Sensing},
volume = {226},
pages = {16-32},
year = {2025},
author = {Shanshan Pan and Runze Zhang and Yilin Liu and Minglun Gong and Hui Huang},
}

@article{pic,
author = {Zhu, Yongning and Bridson, Robert},
title = {Animating sand as a fluid},
year = {2005},
volume = {24},
number = {3},
journal = TOG,
month = jul,
pages = {965–972},
numpages = {8},
}

@inproceedings{cgal,
  title={CGAL: The computational geometry algorithms library},
  author={Fabri, Andreas and Pion, Sylvain},
  booktitle={Proceedings of the 17th ACM SIGSPATIAL international conference on advances in geographic information systems},
  pages={538--539},
  year={2009}
}

@article{mantaflow,
  title={MantaFlow},
  author={Thuerey, Nils and Pfaff, Tobias},
  journal={URL: http://mantaflow. com},
  year={2018}
}

@MISC{eigen,
  author = {Ga\"{e}l Guennebaud and Beno\^{i}t Jacob and others},
  title = {Eigen},
  howpublished = {https://libeigen.gitlab.io},
  year = {2010}
 }

@article{vembook,
  title={The virtual element method},
   volume={32},
    journal={Acta Numerica},
     author={Beirão Da Veiga, Lourenço and Brezzi, Franco and Marini, L. Donatella and Russo, Alessandro}, 
     year={2023}, 
     pages={123–202}
}

@article{extended_cut_cell,
author = {Chen, Yi-Lu and Meier, Jonathan and Solenthaler, Barbara and Azevedo, Vinicius C.},
title = {An extended cut-cell method for sub-grid liquids tracking with surface tension},
year = {2020},
volume = {39},
number = {6},
journal = TOG,
month = nov,
articleno = {169},
numpages = {13},
}

@article{Phase-Field-FLIP,
author = {Braun, Bernhard and Bender, Jan and Thuerey, Nils},
title = {Adaptive Phase-Field-FLIP for Very Large Scale Two-Phase Fluid Simulation},
year = {2025},
volume = {44},
number = {4},
journal = TOG,
month = jul,
articleno = {42},
numpages = {23},
}

@article{c_noc,
    author = {Cangiani, Andrea and Manzini, Gianmarco and Sutton, Oliver J.},
    title = {Conforming and nonconforming virtual element methods for elliptic problems},
    journal = {IMA Journal of Numerical Analysis},
    volume = {37},
    number = {3},
    pages = {1317-1354},
    year = {2016},
    month = {08},
}
\appendix
\section{Construction of Polynomial Basis}
\label{appendix:polynomial}
Following standard choice~\cite{vemtheory}, we use a basis of scaled monomials for all polynomial spaces, defined as
\begin{equation}
     \label{eq:monomials}
m_{\boldsymbol{\alpha}}(\mathbf{x}) := 
\left(\frac{x-x_E}{h_E}\right)^{\alpha_1}
\left(\frac{y-y_E}{h_E}\right)^{\alpha_2}
\left(\frac{z-z_E}{h_E}\right)^{\alpha_3},
\qquad |\boldsymbol{\alpha}| \le k.
\end{equation}
Here, $\mathbf{x}_E$ denotes the centroid of element $E$, and $h_E$ is the diameter, defined as the maximum distance between any two vertices of $E$. The mapping between $\alpha$ and $\boldsymbol{\alpha}=(\alpha_1,\alpha_2,\alpha_3)$ is defined as
\[
1 \leftrightarrow (0,0,0),\quad
2 \leftrightarrow (1,0,0),\quad
3 \leftrightarrow (0,1,0),\quad
4 \leftrightarrow (0,0,1),\ \ldots
\]

\section{Diffusion Analysis of Naive Prolongation Operators}
\label{appendix:diffusion_analyze}
To analyze and mitigate the diffusion effect of naive prolongation operator, the following sets are defined
for a given prolongation matrix $P$:
\begin{equation}
\begin{aligned}
\mathcal{I}(c) &:= \{\, i \mid P(c,i) \neq 0 \,\},\\
\mathcal{I}'(i) &:= \{\, c \mid P^{\mathsf{T}}(c,i) \neq 0 \,\},\\
\mathrm{Link}(i) &:= \{\, j \mid A_f(i,j) \neq 0 \,\}.
\end{aligned}
\end{equation}
Here, $c$ denotes a coarse-level DOF index, while $i$ and $j$ denote fine-level
DOF indices.
The set $\mathcal{I}(c)$ collects fine-level DOFs influenced by the coarse DOF
$c$, $\mathcal{I}'(i)$ collects coarse-level DOFs contributing to the fine DOF
$i$, and $\mathrm{Link}(i)$ encodes the connectivity induced by the fine-level
system matrix $A_f$.
It then follows that a coarse-level matrix entry $A_c(i,j)\neq 0$ if and only if
\[
j \in \bigcup_{k\in \mathcal{I}(c_i)} \mathcal{I}'\!\bigl(\mathrm{Link}(k)\bigr).
\]
\noindent
This implies that, for the naive prolongation operator, an entry $A_c(i,j)$ becomes
nonzero whenever the corresponding DOFs $i$ and $j$ are associated with
polyhedral elements that are $k$-ring neighbors at level $\ell-k$.
Such diffusion leads to a rapid growth of nonzero entries on coarse levels.

\section{Quality Indicator for VEM Cells}
\label{sec:quality_indicator}
The first term,
\begin{equation}
\rho^{3\mathrm{D}}_{1}(E)
=
\frac{|k(E)|}{|E|}
\prod_{f \in \partial E} \frac{|k(f)|}{|f|},
\end{equation}
quantifies the alignment between the cell geometry and its kernel,
penalizing degenerate or strongly non–star-shaped elements.
The kernel of a element $E$ consists of all interior points
from which $E$ is entirely visible.
The second term accounts for local geometric scale variations:
\begin{equation}
\rho^{3\mathrm{D}}_{2}(E)
=
\frac{
\min\!\left(\sqrt[3]{|E|},\; \min\limits_{f \in \partial E} h_f\right)
}{h_E}
+
\frac{1}{\# f_E}
\sum_{f \in \partial E}
\frac{
\min\!\left(\sqrt[3]{|f|},\; \min\limits_{e \in \partial f} |e|\right)
}{h_f},
\end{equation}
\noindent
which measures the consistency between characteristic length scales across
volumetric, face, and edge elements, thereby penalizing strong geometric
imbalance within a cell.
Finally, the third term captures topological complexity:
\begin{equation}
\rho^{3\mathrm{D}}_{3}(E)
=
\frac{4}{\# f_E}
+
\frac{1}{\# f_E}
\sum_{f \in \partial E}
\frac{3}{\# e_f},
\end{equation}
penalizing cells with an excessive number of faces or faces with many edges. For all three terms above,
$\#(\cdot)$ denotes the cardinality of the corresponding set,
and $|\cdot|$ denotes the length, area, or volume of the associated geometric
entity, as appropriate.

\end{document}